\documentclass[particles,article,pdftex,moreauthors]{Definitions/mdpi}

\usepackage{bm}

\firstpage{1} 
\pubvolume{1}
\issuenum{1}
\articlenumber{0}
\pubyear{2026}
\copyrightyear{2026}
\datereceived{ } 
\daterevised{ } 
\dateaccepted{ } 
\datepublished{ } 

\preto{\abstractkeywords}{\nolinenumbers}

\Title{Beaming of polarized radiation in subcritical X-ray pulsars}

\newcommand{\beq}{\begin{equation}}
\newcommand{\dd}{\mathrm{d}}
\newcommand{\Ece}{E_\mathrm{cyc}}

\newcommand{\eeq}{\end{equation}}

\newcommand{\mel}{m_\mathrm{e}}

\newcommand{\nel}{n_\mathrm{e}}

\newcommand{\omce}{\omega_\mathrm{cyc}}
\newcommand{\ompe}{\omega_{\mathrm{pe}}}

\Author{Ivan D. Markozov $^{1,2}$\orcidA{}, Alexander Y. Potekhin $^{1,2,}$*\orcidB{}, Alexander D. Kaminker $^{1}$\orcidC{} and Alexander A. Mushtukov $^{3,4}$\orcidD{}}

\AuthorNames{Ivan D. Markozov, Alexander Y. Potekhin, Alexander D. Kaminker and
Alexander A. Mushtukov}

\address{%
$^{1}$ \quad Ioffe Institute, Politekhnicheskaya 26, St Petersburg 194021, Russia\\
$^{2}$ \quad Space Research Institute RAS, 84/32 Profsoyuznaya str., Moscow 117997, Russia\\
$^{3}$ \quad Mullard Space Science Laboratory, University College London, Holmbury St. Mary, Surrey RH5 6NT, UK\\
$^{4}$ \quad Astrophysics, Department of Physics, University of Oxford, Denys Wilkinson Building, Keble Road, Oxford OX1 3RH, UK}

\corres{Correspondence: palex@astro.ioffe.ru}

\abstract{
Radiation of X-ray pulsars is powered by accretion on the neutron star surface from a binary companion under the influence of a strong magnetic field. We study beaming of this radiation in the case of subcritical X-ray pulsars, where it is formed in the accretion channel close to the neutron star surface.
We solve equations of the hydrodynamics and radiative transfer of two coupled polarization modes in the accretion channel numerically, taking into account resonant Compton scattering and vacuum polarization. The beaming patterns are obtained for different accretion rates, photon energies and polarizations, and for different models of the neutron star surface radiation.
The calculated beaming patterns are converted into 
{light curves for both the intensity and polarization, taking into account}
 the effects of General Relativity.
These beaming patterns and light curves are found to be strongly affected by the resonant Compton scattering for photon energies comparable with the electron cyclotron energy.
{In particular, the angular redistribution of radiation near the cyclotron resonance 
may reduce the light-curve modulation amplitude, 
which is consistent with observational indications of a suppressed pulsed fraction at these energies.}
}


\begin{document}

\section{Introduction}
\label{sec:Intro}

X-ray pulsars (XRPs) are strongly magnetized accreting neutron stars
(NSs) in close binary systems (for a recent review, see
\cite{MushtukovTsygankov24}). In such systems, matter from the
companion star falls onto the NS surface through an accretion channel
along the field lines around magnetic poles. 
The kinetic energy of the
accreting matter that hits the NS surface is mostly converted into
X-ray radiation. Because of a misalignment of the magnetic and rotation
axes of a NS, a remote observer detects a pulsating signal.

A typical field strength at the surface of an XRP is
$B\sim10^{12}-10^{13}$~G.  
In such strong fields, the electron cyclotron energy
\beq
   \Ece = \hbar\omce = \frac{\hbar eB}{m_\mathrm{e} c} =
11.577\,B_{12}\mbox{~keV}
\label{Ecyc}
\eeq
reaches tens keV, causing distinct quantum-mechanical effects.
In \cref{Ecyc},  $\hbar$ is the reduced Planck constant, 
$c$ the speed of light,
$e$ the elementary charge,
$\mel$ the electron mass, $B_{12} \equiv B/10^{12}$~G and
$\omce$ is the electron cyclotron frequency.
{The mean free time of electrons is much larger than $\omce^{-1}$ at typical plasma densities in the accretion channel, which effectively constrains the plasma motion to one dimension along $\bm{B}$.}

Observed X-ray luminosities of the XRPs
{vary}
over 
{several}
orders of magnitude, from 
{$L_\mathrm{X}\lesssim 10^{33}$}
to 
$L_\mathrm{X}\gtrsim 10^{41}$ erg s$^{-1}$ \cite{Yang_17,MushtukovTsygankov24}
(with the reservation that the apparent super-Eddington luminosities of $L_\mathrm{X} \gtrsim 10^{38}$ erg s$^{-1}$ in the ultraluminous sources \cite{Bachetti_14,Israel_17}
may be overestimated because of collimation \cite{KingLasota16,KingKW17,LasotaKing23}, which is a hypothesis under debate
-- see, e.g., Ref.~\cite{Mushtukov_23} and references therein).
At low luminosities $L_\mathrm{X} \lesssim 10^{34}$ erg s$^{-1}$, the XRPs may undergo a slow accretion from a cold disk \cite{Tsygankov_17,2019A&A...621A.134T} or no accretion due to the propeller effect
{\cite{1975A&A....39..185I}}.
Cyclotron absorption lines have been observed in the spectra of nearly 40 XRPs, sometimes with one or a
few harmonics \cite{Staubert_19}. 
Recently, due to the Imaging X-ray Polarimetry Explorer (IXPE) \cite{Weisskopf_22}, it has become
possible to measure the polarization of the X-ray radiation of the XRPs in the band of 2--8 keV 
{(see, e.g., \cite{Heyl22,Doroshenko_22}}.

The bolometric luminosity measured by a distant observer is related to the mass accretion rate $\dot{M}$ onto a NS with mass $M$ and radius $R_\mathrm{NS}$
as
\cite{Mitra98,Meisel_18}
 \beq
  L = 
  \frac{z_\mathrm{g}\dot{M}c^2}{(1+z_\mathrm{g})^2}
  =
  10^{36}\,\frac{1.327}{1+z_\mathrm{g}/2}\,
  \frac{M/M_\odot}{R_6}\,
  \dot{M}_{16}
  \mbox{ erg s}^{-1},
\eeq
where 
$z_\mathrm{g} = (1-2GM/c^2 R_\mathrm{NS})^{-1/2} - 1$ is the gravitational redshift,
$G$ is the Newtonian constant of gravitation,
$M_\odot$ the solar mass,
$R_6 = R_\mathrm{NS}/10^6$~cm
and $\dot{M}_{16} = \dot{M}/10^{16}$ g s$^{-1}$.
{Here $\dot{M}$ is measured in the local reference frame at the NS surface;
the accretion rate measured by a remote observer is $\dot{M}^\infty = \dot{M}/(1+z_\mathrm{g})$.}
There is a critical value of luminosity ($L_\mathrm{crit}\sim10^{36}$ erg s$^{-1}$ for a typical XRP), at which radiation pressure can stop plasma motion \cite{Davidson73}. 
XRPs with
{higher luminosities}
are called supercritical, while XRPs with lower
{luminosities}
are called subcritical. In the supercritical XRPs, a radiation-dominated shock is expected
{in the accretion channel} \cite{BaskoSunyaev76}. 
The observed radiation of the supercritical XRPs originates
{from the accretion column, made of a slow moving dense hot plasma below the shock front.}

For the subcritical XRPs, two fundamentally different accretion regimes have been discussed.
One of them is determined by Coulomb braking of the
{fast moving}
plasma in a relatively thin layer near the NS surface 
{\cite{1969SvA....13..175Z}}.
This process leads to
formation of
{heated}
atmosphere regions around the magnetic poles, from which the
X-ray radiation emerges
{(e.g., \cite{Sokolova-Lapa_21})}. 
In the other regime
\cite{ShapiroSalpeter75,LangerRappaport82,BykovKrassilchtchikov04},
a collisionless shock wave appears, which plays the crucial role in the accretion channel structure and radiation formation. 
In the present paper we will focus on the subcritical XRPs in the shockless regime.

In a previous work \cite{MarkozovKP23} 
we constructed a model of the accretion channel
and calculated polarized radiation transfer
in the electron-proton plasma approximation for the dielectric tensor.
{The model included the resonant scattering of radiation in a strong magnetic field, but neglected the damping of the electron cyclotron resonance, the}
ion cyclotron motion and vacuum polarization effects. 
Besides, we considered 
only the case of a completely filled accretion 
channel, which can probably be formed under the conditions of spherical accretion.
A more realistic geometry of disc accretion favors formation of a hollow channel, 
only partially filled with plasma \cite{BaskoSunyaev76}.
A detailed study of the plasma hydrodynamics and polarized radiation transfer in the accretion channels of the subcritical XRPs with account of the vacuum polarization in both accretion channel geometries will be presented elsewhere \cite{Markozov_26}.
Here we will focus on
the influence of radiative transfer in the hollow accretion channel 
on the beaming of the outgoing radiation. As well as in the above-cited papers, the 
plasma is assumed to be composed of fully ionized hydrogen.

\section{Physics input and numerical methods}
\label{sec:input}

Propagation of electromagnetic waves in magnetized plasmas has been
comprehensively described in the monograph by \citet{Ginzburg70}. 
At circular frequencies $\omega$, which lie sufficiently far from resonances and 
are much larger than the electron plasma
frequency $\ompe=\left({4\pi e^2 \nel / \mel} \right)^{1/2}$, where $\nel$ is the electron number density, radiation
propagates in the form of
two normal modes.
The normal mode whose electric vector oscillates along an ellipse with a major axis lying in the plane formed by the photon wave vector $\bm{k}$ and magnetic field vector $\bm{B}$ is called ordinary (O-mode), and the normal mode whose polarization ellipse has a major axis
{perpendicular}
to this plane is
called extraordinary (X-mode).
{The two modes}
have different absorption and scattering cross sections, which
depend on the angle $\theta_B$ between $\bm{k}$ and $\bm{B}$,
{and}
interact with
{each other}
through scattering. 

Gnedin and Pavlov~\cite{GnedinPavlovModes} formulated the radiative
transfer problem in terms of the normal modes and specified the
{applicability}
conditions of
{such description}.
 They showed that in the conditions typical for strongly
magnetized NSs a strong
Faraday depolarization occurs at most photon energies $E=\hbar\omega$ (except narrow frequency ranges near resonances), which allows one to consider specific intensities of the two
normal modes instead of the four components of the Stokes vector.

The influence of the vacuum polarization on the NS radiation was
first evaluated in \cite{Novick_77,GnedinPS78} and
studied in detail in subsequent works,
summarized by Pavlov and Gnedin \cite{PavlovGnedin84}.
A further progress was made by Lai et al. \cite{LaiHo02,HoLai03,LaiHo03,vanAdelsbergLai06},
who used a more general formalism and studied the effect of normal mode conversion for radiation propagating in an inhomogeneous plasma
across
{so called}
vacuum resonance.
{At this resonance,}
the vacuum and plasma polarization effects 
become equally important and the normal mode approximation is no longer valid.
If the density variation is sufficiently gentle, 
the polarization of a photon evolves adiabatically, preserving 
the electric vector rotation direction,
but changing the direction of the major axis of the rotation ellipse relative to the $\bm{k}-\bm{B}$ plane, which means
{the}
conversion between the O-mode and X-mode. This occurs at a plasma density 
$
  \rho_\mathrm{vac} \sim 10^{-4}B_{12}^2(E/\mbox{1~keV})^2\text{ g cm}^{-3}
$.
Here we follow the general formalism of Lai and Ho \cite{LaiHo03}, but neglect the mode conversion, because the subcritical accretion channel 
is almost transparent to radiation at 
{those relatively low}
photon energies,
{at which the plasma density $\rho$ in the channel can reach $\rho_\mathrm{vac}$}.
The non-trivial effects of Compton scattering in the accretion channel occur at higher $E$, where the X- and O-mode polarization vectors are almost linear, being determined mostly by the vacuum polarization effects.

Compton scattering in a strong magnetic field differs significantly from the field-free one. We treat it by analogy with our previous papers \cite{Mushtukov_22,MarkozovKP23}, but with certain improvements and corrections, as described in detail in \cite{Markozov_26}. 
{We}
assume that an electron  occupies the ground Landau level before and after scattering, which is generally a good approximation for the XRPs because of the quick radiative decay of the excited Landau states \cite{Meszaros_book}. The differential
cross sections of photon scattering on an electron in a strong magnetic field is adopted from Ref.~\cite{Herold79}, where it is given for an electron at rest and for the linear polarizations of the incoming and scattered photons.  We
{transform the cross sections}
to the elliptical normal-mode
{polarization basis and average them}
over the relativistic thermal distribution of electron momenta parallel to $\bm{B}$ (the one-dimensional Maxwell--J\"uttner distribution -- see, e.g., Appendix A in Ref.~\cite{Mushtukov_22}),
{using the}
Lorentz transformations of
energies $E$ and angles $\theta_B$ for the incoming and scattered photons
{of the cross sections}. These cross sections are then employed in the numerical solutions of equations of radiation hydrodynamics for the one-dimensional plasma
motion along the magnetic field lines and stationary radiative transfer equations for the two coupled normal modes. The stationary radiative transfer approximation is applicable because the typical time of photon propagation inside the accretion channel is short on
the hydrodynamic time scale. In solving the system of radiation hydrodynamics equations we  assume that electrons and protons have the same temperature $T$ and apply an ideal gas equation of state (EoS), which is a good approximation at the XRP conditions, because $T\sim$ a few keV is much
{higher}
than the hydrogen binding energy at $B\lesssim10^{13}$~G. This plasma EoS is supplemented by the EoS of ideal photon gas, assuming the local thermodynamic equilibrium. However, as mentioned in Ref.~\cite{MarkozovKP23}, the exact EoS form
is unimportant for our model calculations, because the accretion flow is mainly controlled by the
{gravity and radiation pressure},
rather than by the
{EoS}.

For a numerical solution of the problem, we use the time discretization and apply the operator  splitting method \cite{LeVeque_2002}.  Each time step is divided into three  substeps.

At the first substep we solve the hydrodynamical  system of equations, neglecting the  radiative terms. To solve this simplified system, we employ the open  library VH-1 (`Virginia Hydrodynamics-1' \cite{ColellaWoodward84}).%
\footnote{\href{http://wonka.physics.ncsu.edu/pub/VH-1/}{http://wonka.physics.ncsu.edu/pub/VH-1/}}
The numerical grid  is organized as a hollow cylinder of the outer radius
$R_\mathrm{c}$, inner radius $R_\mathrm{c}-d_\mathrm{w}$ ($d_\mathrm{w}$ being the thickness of the cylindrical walls) and height $H$,  divided into a sufficient number of equal slices. We neglect variations  of all hydrodynamic variables
along  the radius of the cylinder,  because we have found previously \cite{MarkozovKP23} that these  quantities do not significantly  vary along the radial coordinate.
The magnetic field and the cylinder axis are perpendicular to the NS surface. In the examples of calculations presented below, we have set $M=1.4\,M_\odot$,
$R_\mathrm{NS}=12$ km. $R_\mathrm{c} = 0.5$~km and $d_\mathrm{w} = 15$~m.  The height $H$ is chosen sufficiently large  to exceed the height of the zone where radiation pressure is significant,  so that the plasma velocity at the upper face of the cylinder
almost coincides with the free-fall velocity.
Nevertheless $H \ll R_\mathrm{NS}$ in the subcritical regime
{\cite{SuleimanovLS07}}.
The magnetic field strength $B=4.319\times10^{12}$~G is 
{chosen to provide the cyclotron energy}
$\Ece=50$~keV, which is typical for the XRPs.

At  the second substep, radiation transfer of the two coupled normal modes and the plasma-radiation energy-momentum exchange are simulated by the Monte Carlo method (see Ref.~\cite{Mushtukov_22} for a detailed description of our Monte Carlo code). To accelerate
these simulations, we use precalculated tables of the total scattering cross sections and cumulative distribution functions of probability of photon scattering into a given direction and given normal mode with initial $E$ and $\theta_B$. 
These tables are based on the differential cross sections and depend on $\rho$,  $T$ and $\Ece$ as parameters; their detailed description is given in Ref.~\cite{Markozov_26}.

At the third substep, the contribution of the radiative terms, which have been neglected at the first substep, is restored. The energy-momentum exchange between the electron-proton plasma and the photons is described by a system of first-order differential equations for time conservation of  mass, energy and momentum densities. These equations are solved by  the explicit Euler method, which has been found appropriate in the case of subcritical pulsars.

We assume that the plasma is in the free-fall state at the top of the accretion channel and set its inflow velocity equal to the free-fall velocity
$v_\mathrm{ff}= 
{\sqrt{2GM/(R_\mathrm{NS}+H)} \approx} \sqrt{2GM/R_\mathrm{NS}}$. 
The plasma density at the top boundary is determined by this velocity,
the mass accretion rate $\dot{M}$ and the
channel cross-section area $2\pi R_\mathrm{c}d_\mathrm{w}$,
{that is $\rho(H) = \dot{M}/2\pi R_\mathrm{c}d_\mathrm{w} v_\mathrm{ff}$.}

\begin{figure}[ht]
\includegraphics[width=.53\columnwidth]{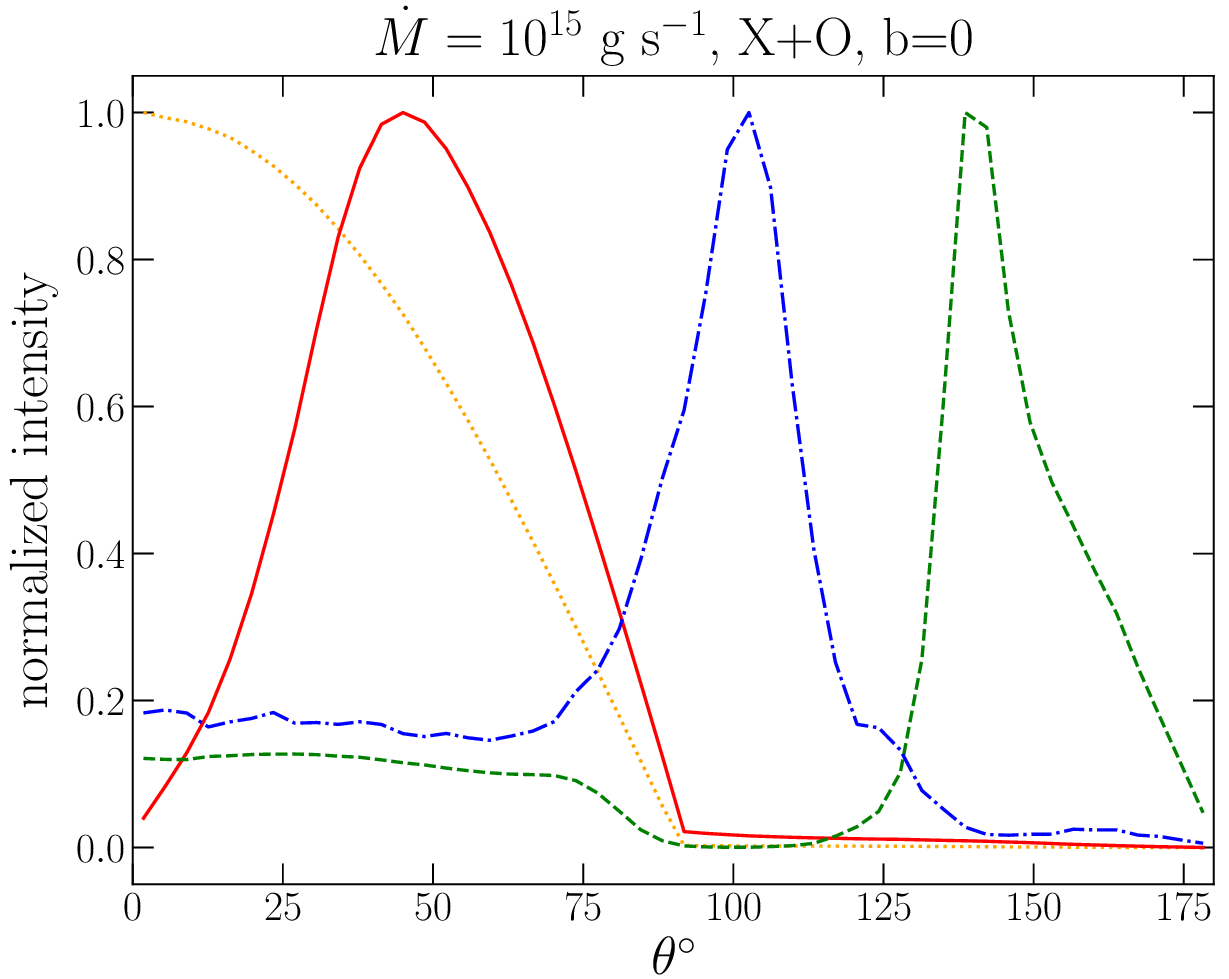}
\hfill
\includegraphics[width=.43\columnwidth]{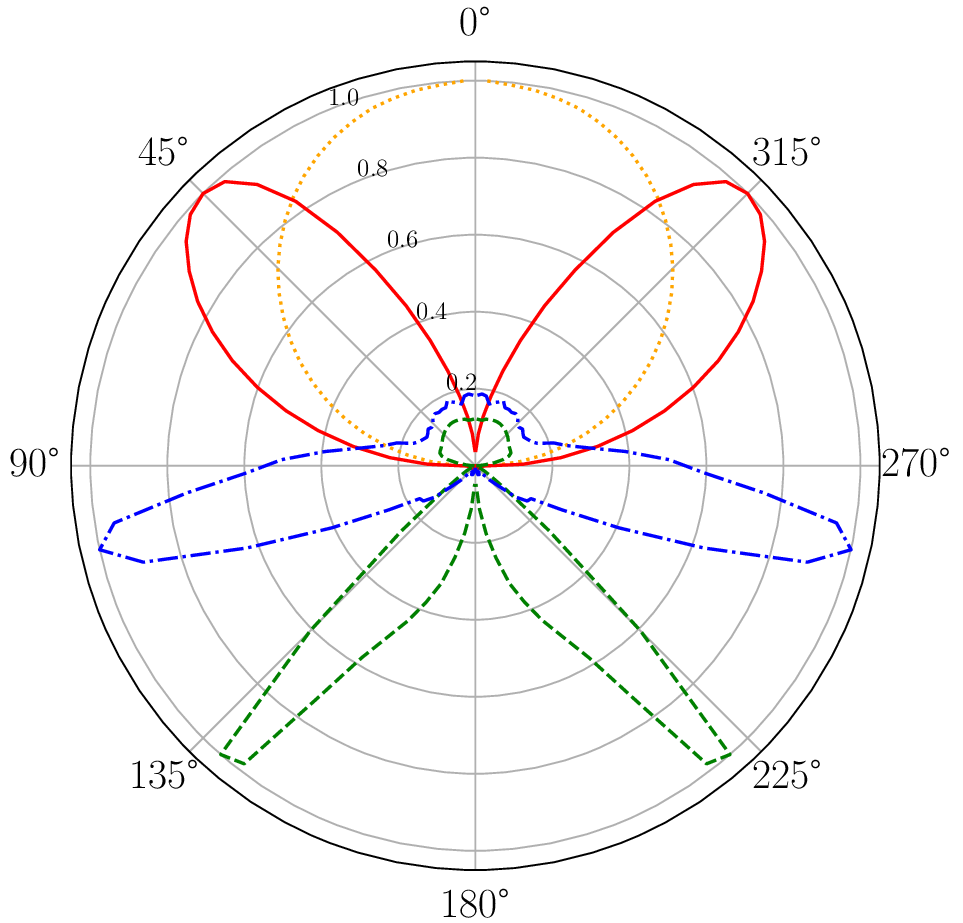}
\caption{Dependence of the normalized radiant intensity $Q_E$ on the polar angle $\theta$ (left panels) and corresponding beam pattern (right panel) at accretion rate $\dot{M} = 10^{15}$ g s$^{-1}$ for photon energies $E = 0.2\Ece$ (dotted orange lines), $0.4\Ece$ (solid red lines), $\Ece$ (dashed green lines) and $1.5\Ece$ (dot-dashed blue lines).
The specific intensity $I_E(\theta)$ at the surface is assumed constant (parameter $b=0$).
Each curve is normalized to its maximum.
\label{fig:gen10}}
\end{figure}   
\unskip
\begin{figure}[ht]
\includegraphics[width=.53\columnwidth]{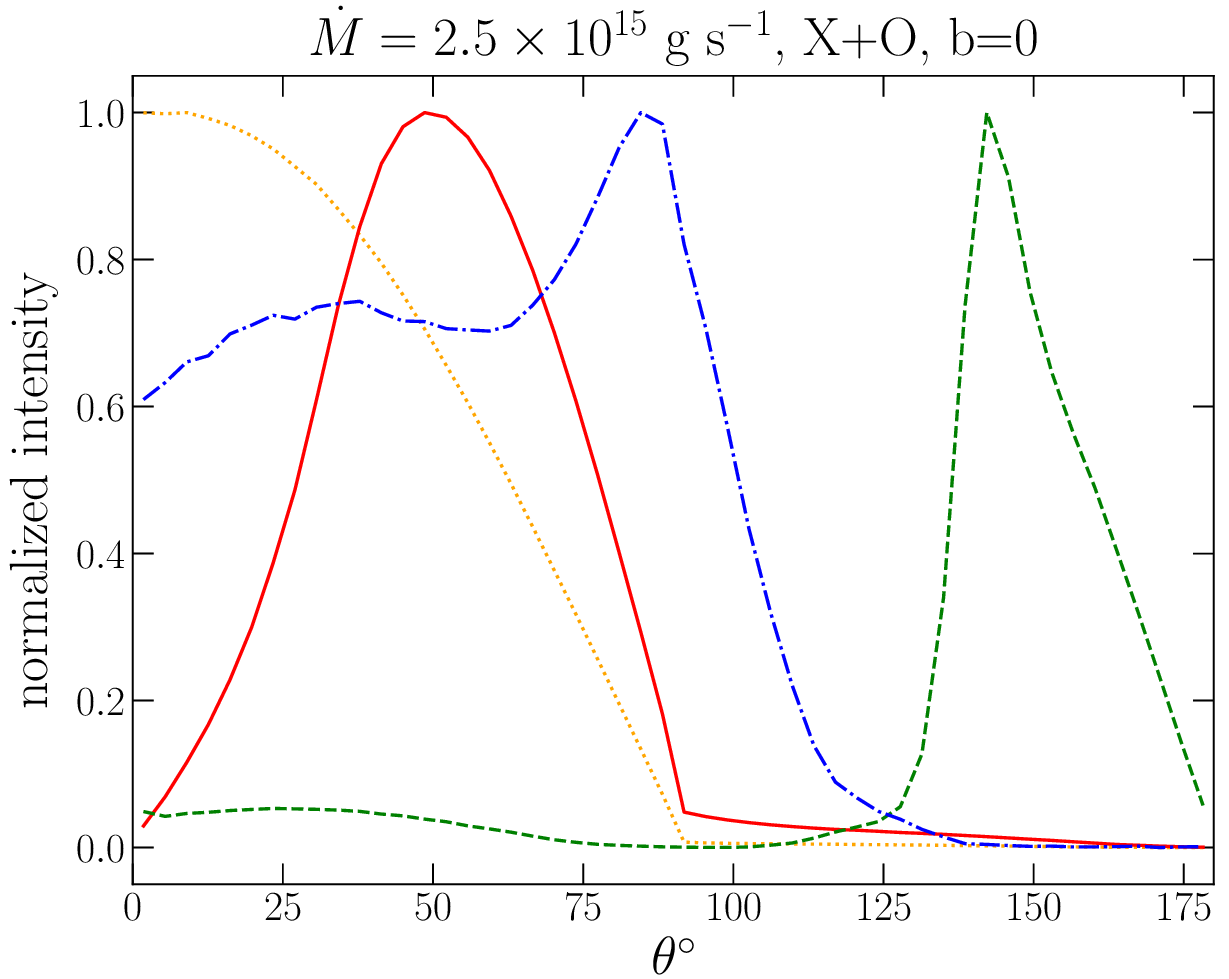}
\hfill
\includegraphics[width=.43\columnwidth]{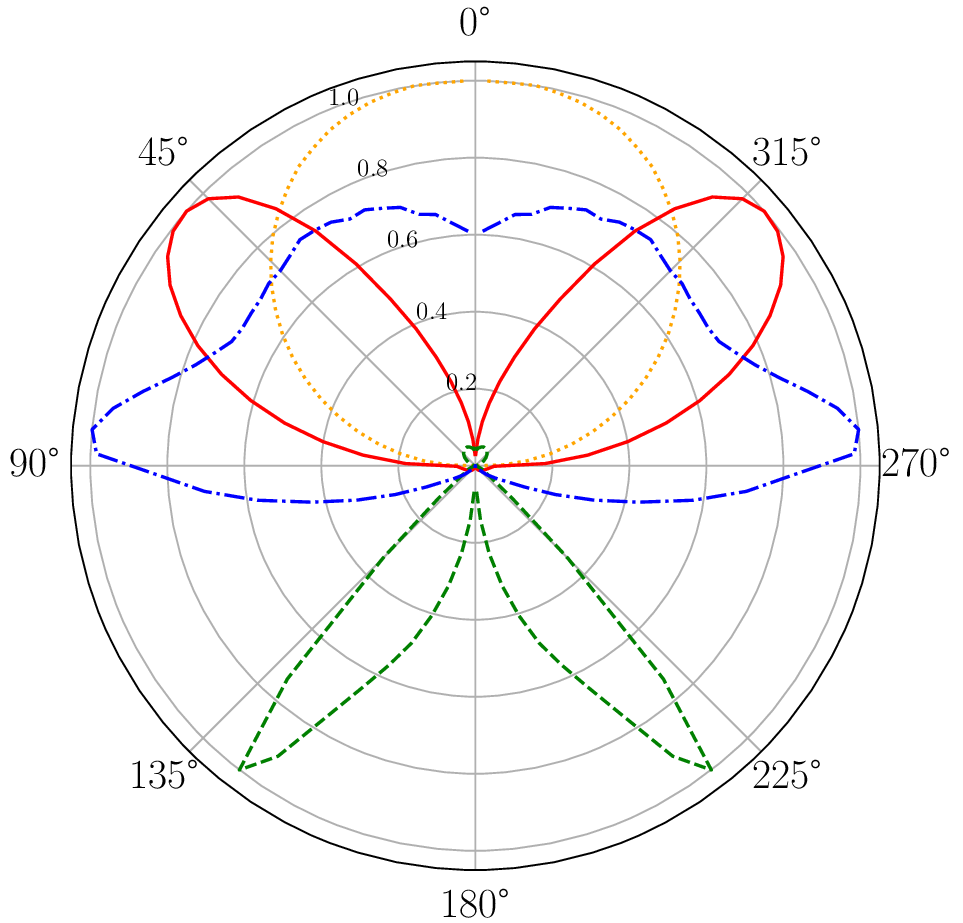}
\caption{Same as in \cref{fig:gen10} but for $\dot{M} = 2.5\times10^{15}$ g s$^{-1}$.
}
\label{fig:gen25}
\end{figure}
\unskip
\begin{figure}
\includegraphics[width=.53\columnwidth]{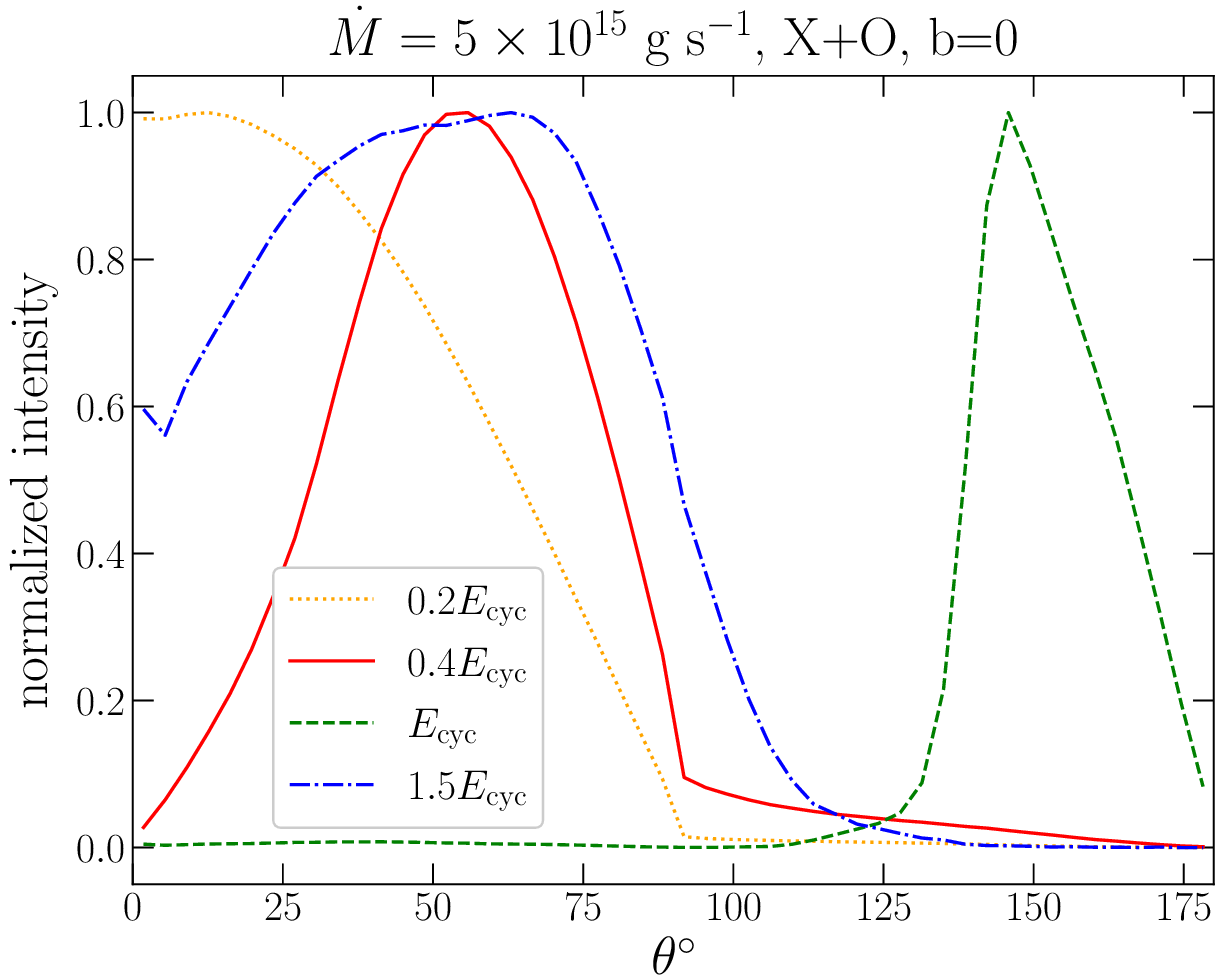}
\hfill
\includegraphics[width=.43\columnwidth]{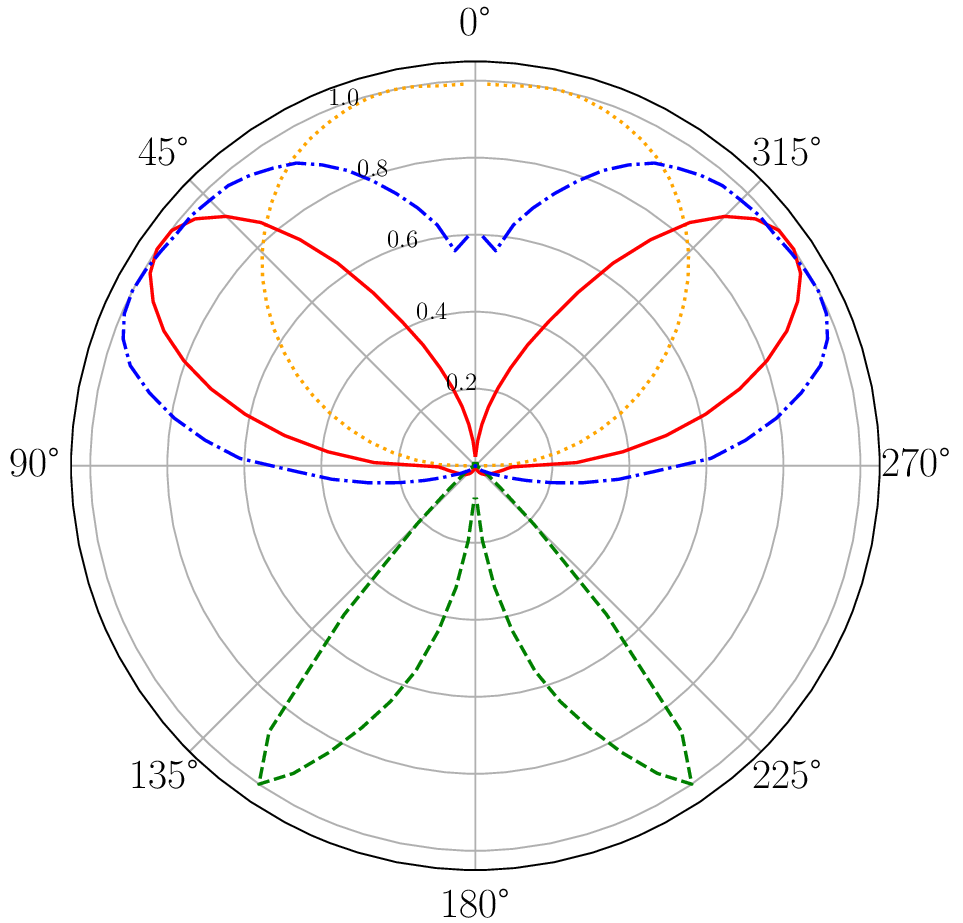}
\caption{Same as in \cref{fig:gen10,fig:gen25} but for $\dot{M} = 5\times10^{15}$ g s$^{-1}$.
\label{fig:gen50}
}
\end{figure}

The boundary conditions at the bottom of the channel are not so obvious.
In our model, the bottom is placed at the NS surface. Therefore the
proper bottom boundary conditions are determined by the physics of the
interaction of the accreting plasma and radiation with the NS surface or
the atmosphere, which is a complex, not completely 
{solved} problem.
In the examples of simulations presented below, we assume that the total (plasma and radiation) energy that enters the NS surface is re-emitted by electromagnetic radiation with a specific intensity $I_E(\theta)$, where $\theta$ is an angle between the wave vector $\bm{k}$ and the NS surface normal.
{We assume that the surface radiation}
has the blackbody spectrum with an uniform temperature determined by the energy balance within the emitting ring, but we allow for possibly non-trivial initial polarization or beaming.
For the polarization, we test the three extreme cases of non-polarized surface radiation (i.e., equal numbers of X- and O-photons of the same energy) and the radiation polarized entirely in either the X-mode or the O-mode.
For the initial beaming, we apply the simplest non-isotropic model
\beq
   I_E(\theta) \propto 1 + b\,\cos\theta,
\label{I_E_b}
\eeq
where $b\geq-1$ is a free parameter.
Then the radiant intensity (the energy per unit solid angle emitted by the hot spot of area $S$ at the NS surface) $Q_E=S_\perp I_E(\theta) = S I_E(\theta) \cos\theta$, where $S_\perp=S\cos\theta$ is the projected transverse hot-spot area,
follows the Lambert law $Q_E\propto\cos\theta$ at $b=0$, while it is suppressed along the normal to the surface at $b < 0$ and more elongated at $b > 0$.

\section{Results}
\label{sec:res}

\subsection{General features of beaming}
\label{sec:beam}

In the left panels of \cref{fig:gen10,fig:gen25,fig:gen50} we present the calculated $\theta$-dependencies of the radiation outgoing from the accretion channel at different photon energies. In the right panels, we show the corresponding beaming patterns in the polar coordinates.
The surface radiation is assumed unpolarized. Each function $Q_E(\theta)$ is normalized to its maximum, in order to focus on the angular dependence. It should be noted that the parts of the curves with $\theta\geq90^\circ$ cannot be observed: since the radius and height of the accretion channel are small compared to the NS radius, the radiation that comes out at these angles goes back to the NS surface and contributes to its heating.

At the lowest energy $E=0.2\Ece$, the optical thickness of the accretion channel is small, therefore the corresponding (dotted) line is close to the initial surface beaming, which in this case was assumed isotropic (parameter $b$ in \cref{I_E_b} is zero).
At all other energies shown in the figure, including the rather low value $E=0.4\Ece$, the beaming of the outgoing radiation is strongly anisotropic, unlike the initial one, which means that this pattern is formed due to the scattering in the accretion channel. 

The fan-like pattern at $E=0.4\Ece$ with strongly suppressed intensity
{around}
the normal to the surface appears due to the large optical depth for the photons propagating at small angles $\theta$ nearly along the cylindrical walls. The wall thickness $d_\mathrm{w}$ is much smaller than the cylinder height, so that the photons with large inclination angles escape easily.

For $E=\Ece$, the resonant Compton scattering diverts most photons back to the NS surface, which explains their concentration at $\theta>90^\circ$.

At a higher energy $E=1.5\Ece$, multiple non-resonant scattering in the channel is substantial. At a low accretion rate $\dot{M}=10^{15}$ g s$^{-1}$ (\cref{fig:gen10}), the beaming pattern still resembles the pattern at the resonance, although a significant (albeit minor) fraction of photons escapes in the upper hemisphere $\theta < 90^\circ$. With increasing $\dot{M}$ (\cref{fig:gen25,fig:gen50}), the plasma density in the channel increases, the mean free path of a photon decreases respectively, and the multiple scattering becomes more important. In these cases
 we observe a broad angular distribution, slightly suppressed 
{around}
the normal.
 
\subsection{Influence of boundary conditions}
\label{sec:boundary}

\begin{figure}[ht]
\includegraphics[width=\columnwidth]{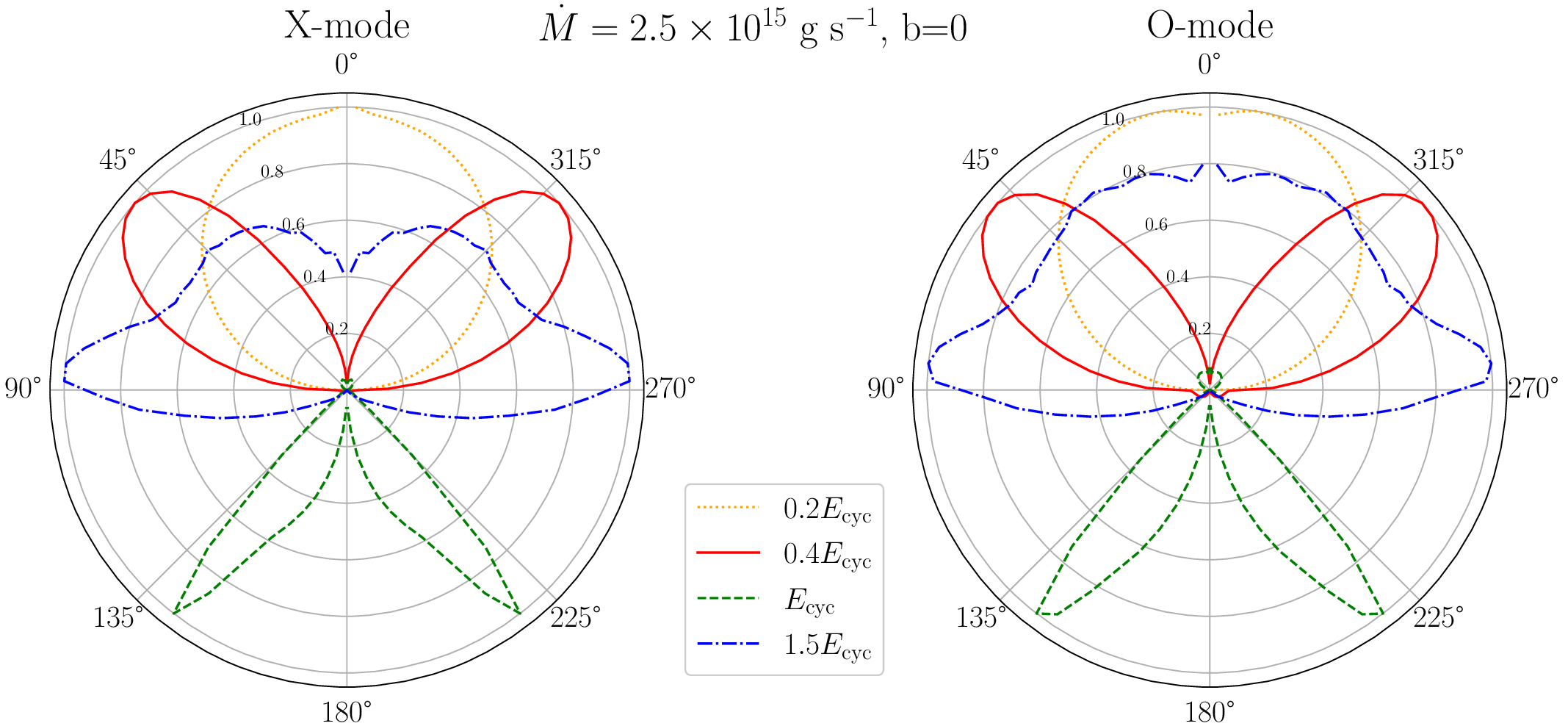}
\caption{Polar diagram of outgoing radiation at $\dot{M}=2.5\times10^{15}$ g s$^{-1}$ (as in
\cref{fig:gen25}) for separate  polarizations of NS surface radiation: incoming X-mode (left panel) and incoming O-mode (right panel).
\label{fig:pol_in}}
\end{figure}

\begin{figure}[ht]
\includegraphics[width=\columnwidth]{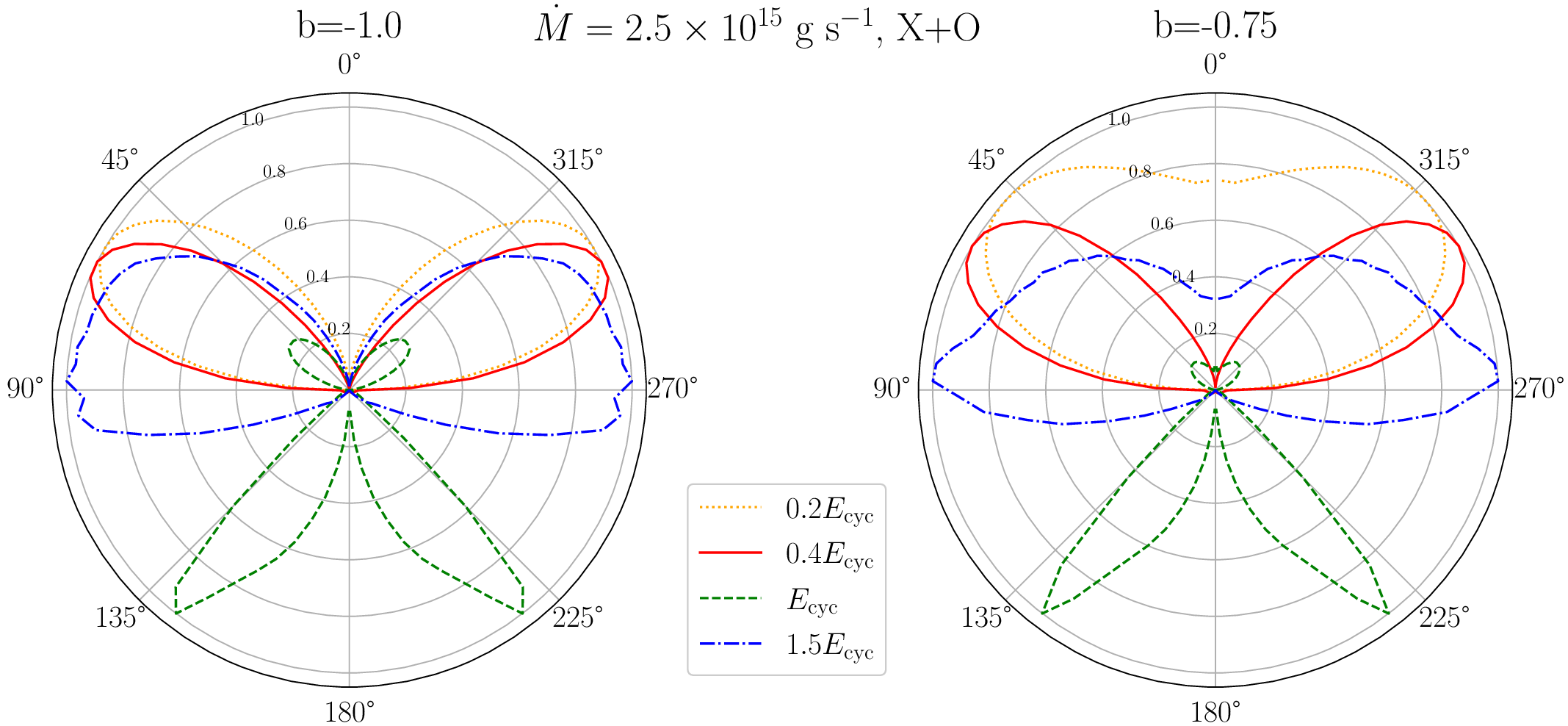}
\caption{Polar diagram of beaming of outgoing radiation at $\dot{M}=2.5\times10^{15}$ g s$^{-1}$ for the unpolarized NS surface radiation, suppressed near the normal to the surface according to \cref{I_E_b} with parameter $b=-1$ (left panel) or $b=-0.75$ (right panel).
\label{fig:beamed}}
\end{figure}

\Cref{fig:pol_in,fig:beamed} show beaming of the outgoing radiation calculated assuming different boundary conditions. 
\Cref{fig:pol_in} demonstrates the case where the radiation that enters the accretion channel from the NS surface is
polarized. The left panel presents the case where the initial polarization is completely in the X-mode and the right panel corresponds to the initial O-mode. Here, the accretion rate $\dot{M}=2.5\times10^{15}$ g s$^{-1}$ is assumed, as in  \cref{fig:gen25}.
We see that this change of boundary conditions affects the beaming noticeably, but not crucially. The beaming patterns at the different energies remain similar to those in \cref{fig:gen25}, although they are somewhat broader in the case where the initial polarization is ordinary compared with the
{opposite case of}
extraordinary initial polarization. The most significant difference occurs at the energy above the cyclotron resonance,
{where}
the broadening of the beaming pattern
{is significantly more anisotropic in the case of the initial X-mode}.

\unskip 

\Cref{fig:beamed} demonstrates beaming of the anisotropic radiation, coming into the bottom of the accretion channel from the NS surface according to \cref{I_E_b} with negative parameter $b$.
{This model}
corresponds to the initial beam pattern suppressed along the normal to the surface, as expected at very low accretion rates \cite{Mushtukov_21b,Sokolova-Lapa_21}. Here the incoming surface radiation is unpolarized and the accretion rate is $\dot{M}=2.5\times10^{15}$ g s$^{-1}$, as in \cref{fig:gen25}. The right panel presents the case of moderate
{initial anisotropy}
with $b=-0.75$ and the left panel shows the extreme case of $b=-1$. 
We see that the scattering in the accretion channel strongly modifies the beaming at not too low photon energies $E\geq0.4\Ece$. Still the beaming pattern retains traces of the initial angular distribution.
In particular, the radiation remains completely suppressed in the normal direction $\theta=0$ in the case of $b=-1$.

\begin{figure}[ht]
\includegraphics[width=\columnwidth]{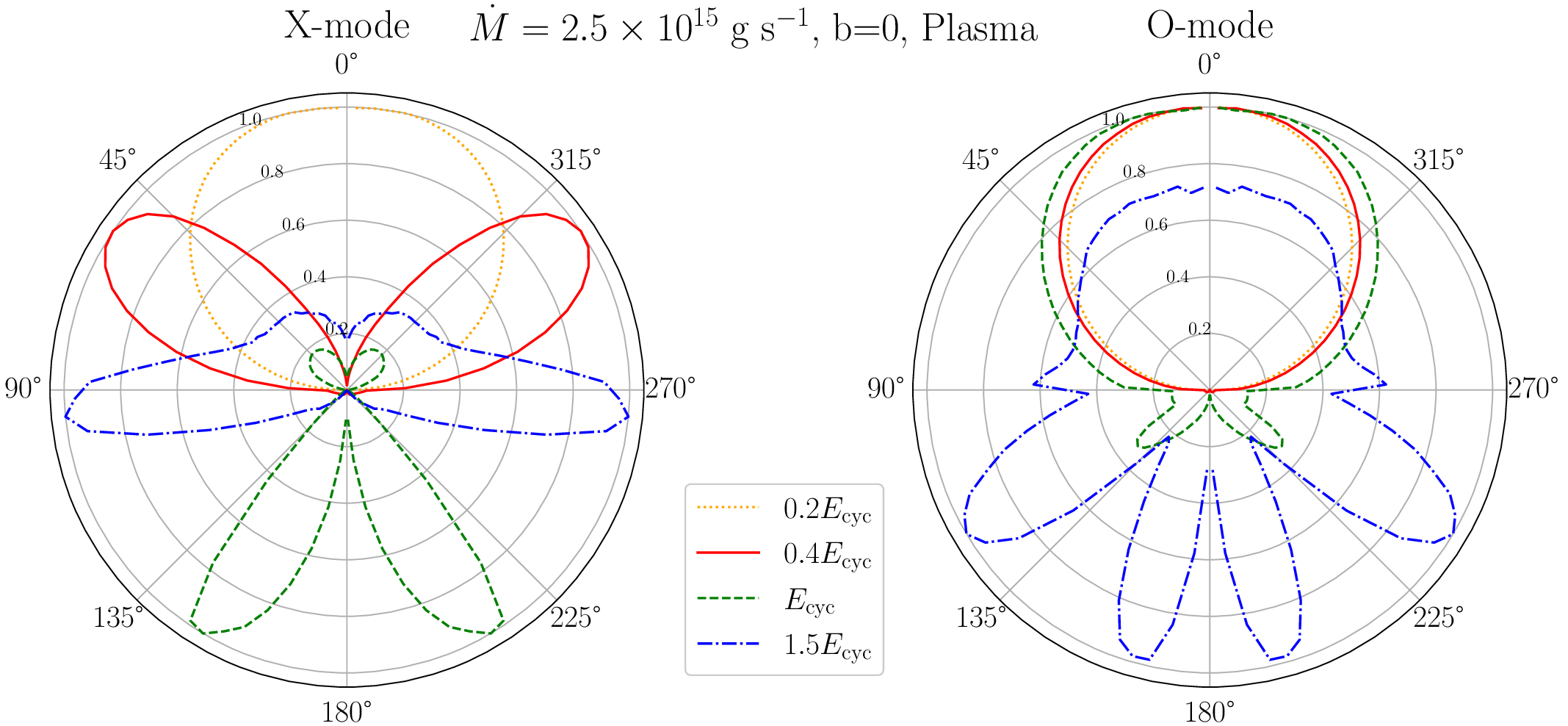}
\caption{Same as in \cref{fig:pol_in}, but in the ``plasma approximation'' where the vacuum polarization effects are neglected.
\label{fig:plasma}}
\end{figure}

\subsection{Importance of vacuum polarization}
\label{sec:vacpol}

In the previous paper \cite{MarkozovKP23} we simulated the radiation outgoing from the accretion channels of subcritical XRPs taking into account the birefringence of strongly magnetized plasma, but neglecting the quantum-electrodynamical effect of vacuum polarization \cite{PavlovGnedin84}. 
{Recently, Sokolova-Lapa et al.~\cite{Sokolova-Lapa_23} argued that this effect should be important for formation of XRP spectra. Our}
new simulations fully take the vacuum polarization into account following the formalism of Ref.~\cite{LaiHo03} (full details of our treatment are given in Ref.~\cite{Markozov_26}).
{We show that} the vacuum polarization effect {is important not only for the spectra, but also for the beaming. It can be} seen from a comparison of \cref{fig:pol_in} with \cref{fig:plasma},
{the latter being different}
from the former one by neglecting the vacuum polarization effects. 
The comparison reveals strong differences in the beaming patterns. A more complex beaming is formed if the vacuum polarization is neglected, in which case the normal mode polarization ellipses are less elongated. The sharpest difference is observed in the case where the NS surface radiation is polarized in the O-mode (compare the right panels of \cref{fig:pol_in,fig:plasma}).

\begin{figure}[ht]
\includegraphics[width=.53\columnwidth]{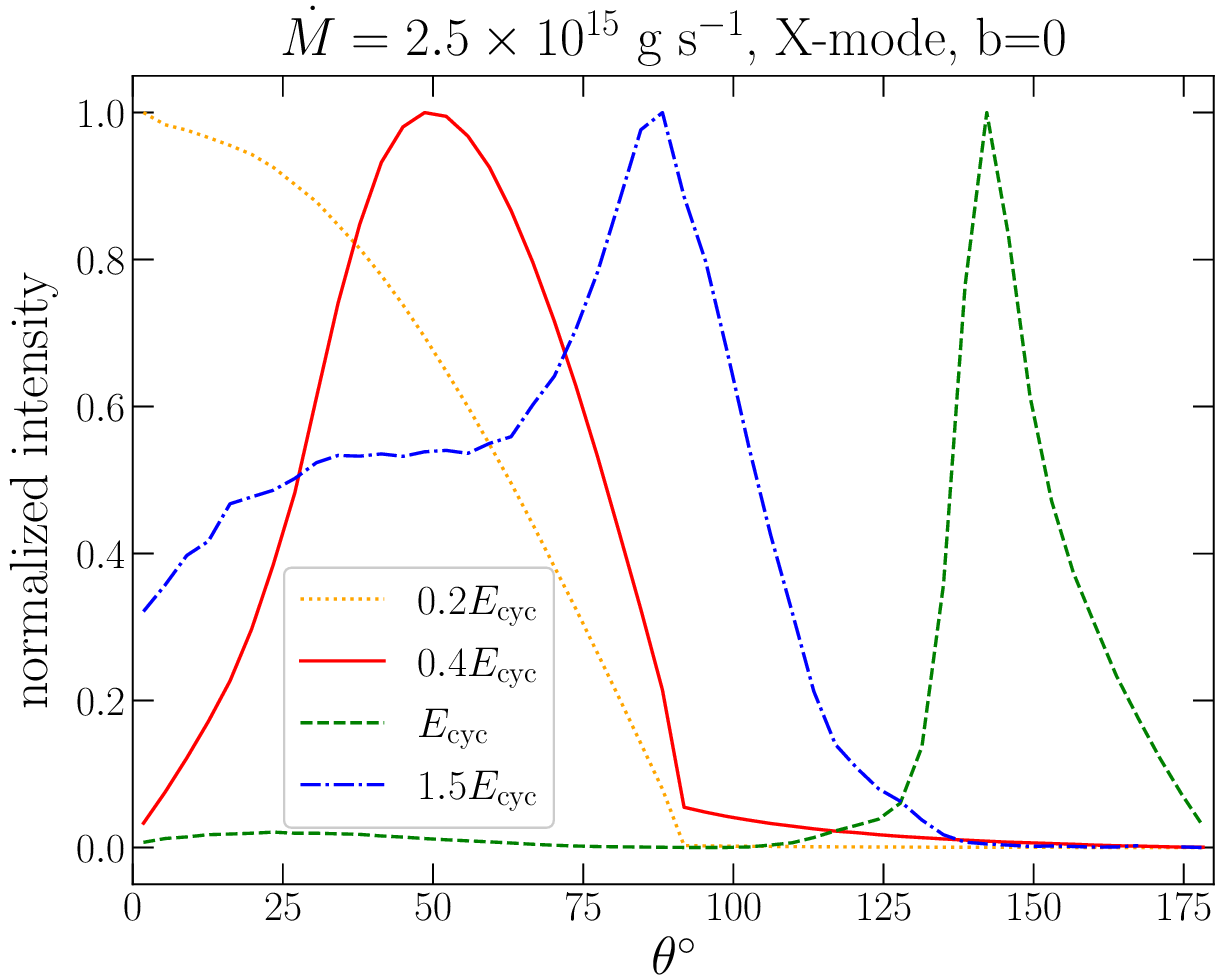}
\hfill
\includegraphics[width=.43\columnwidth]{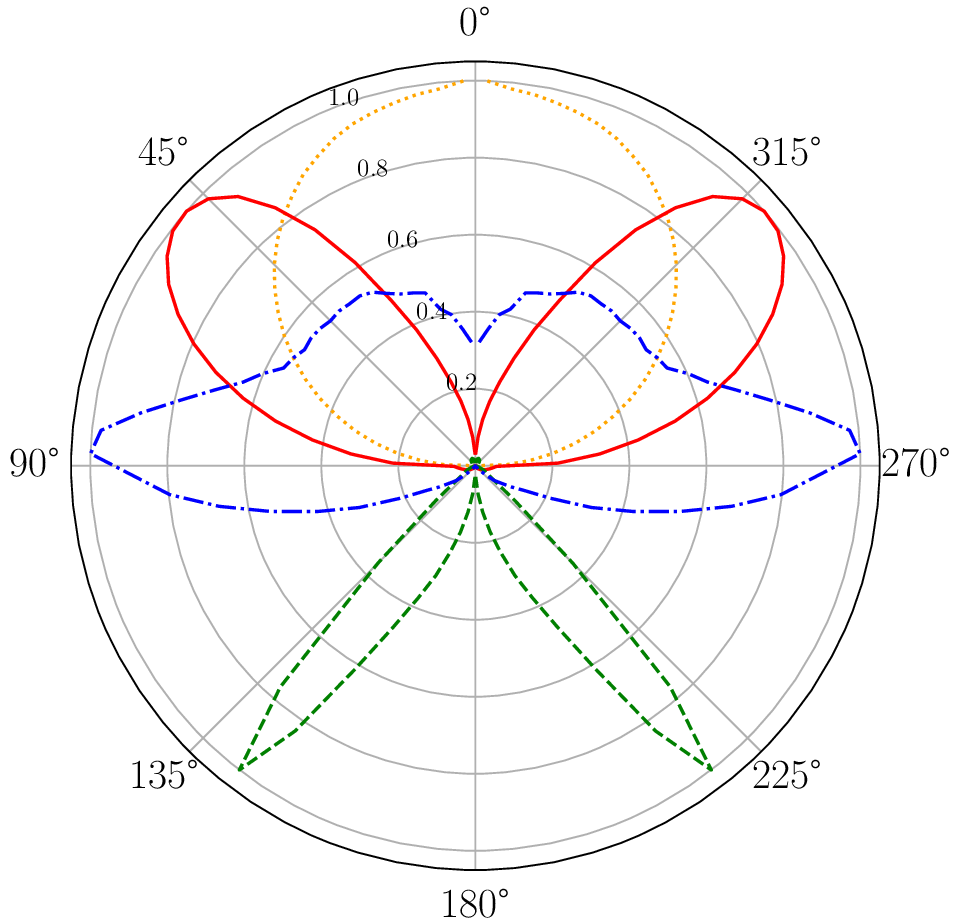}
\\
\includegraphics[width=.53\columnwidth]{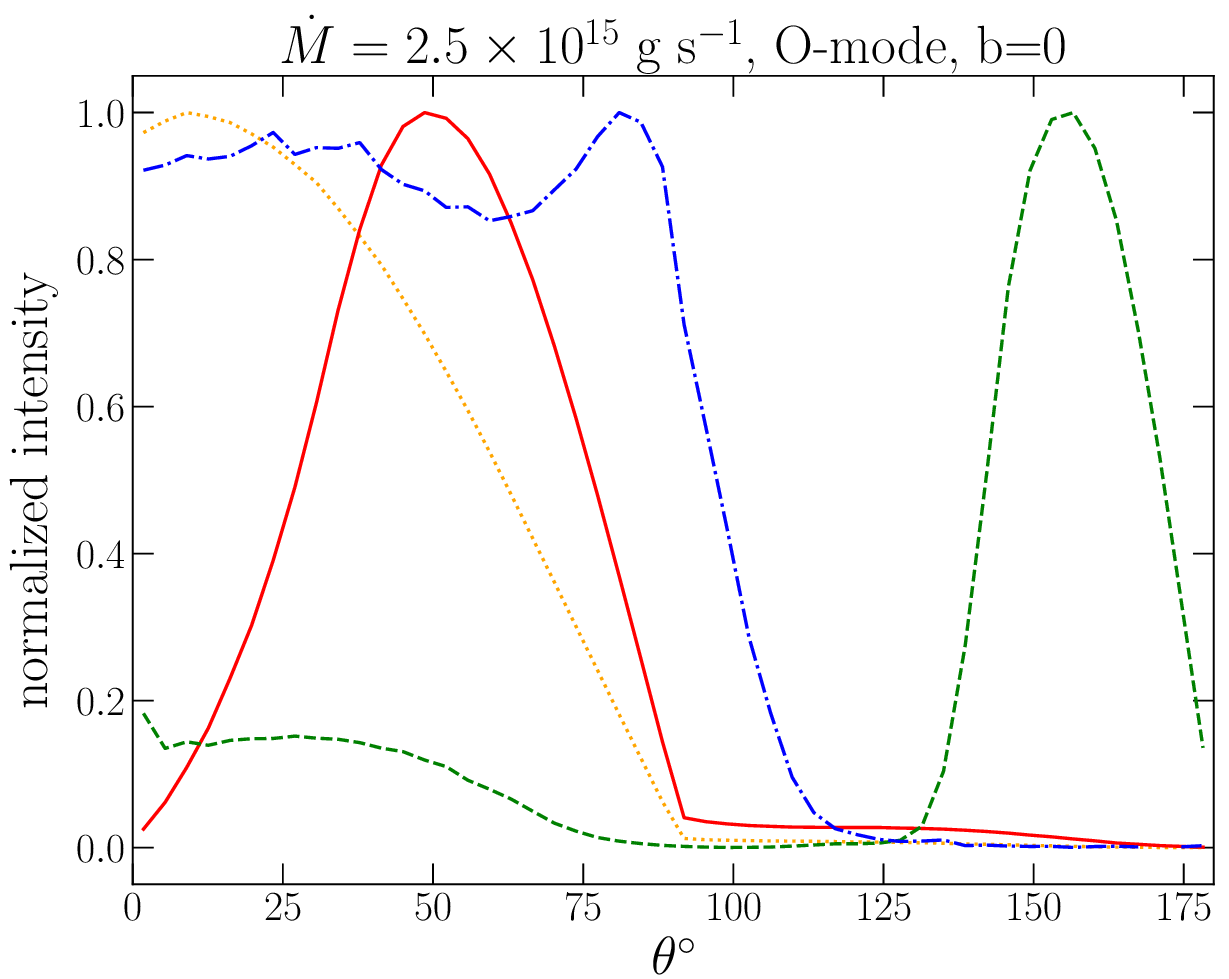}
\hfill
\includegraphics[width=.43\columnwidth]{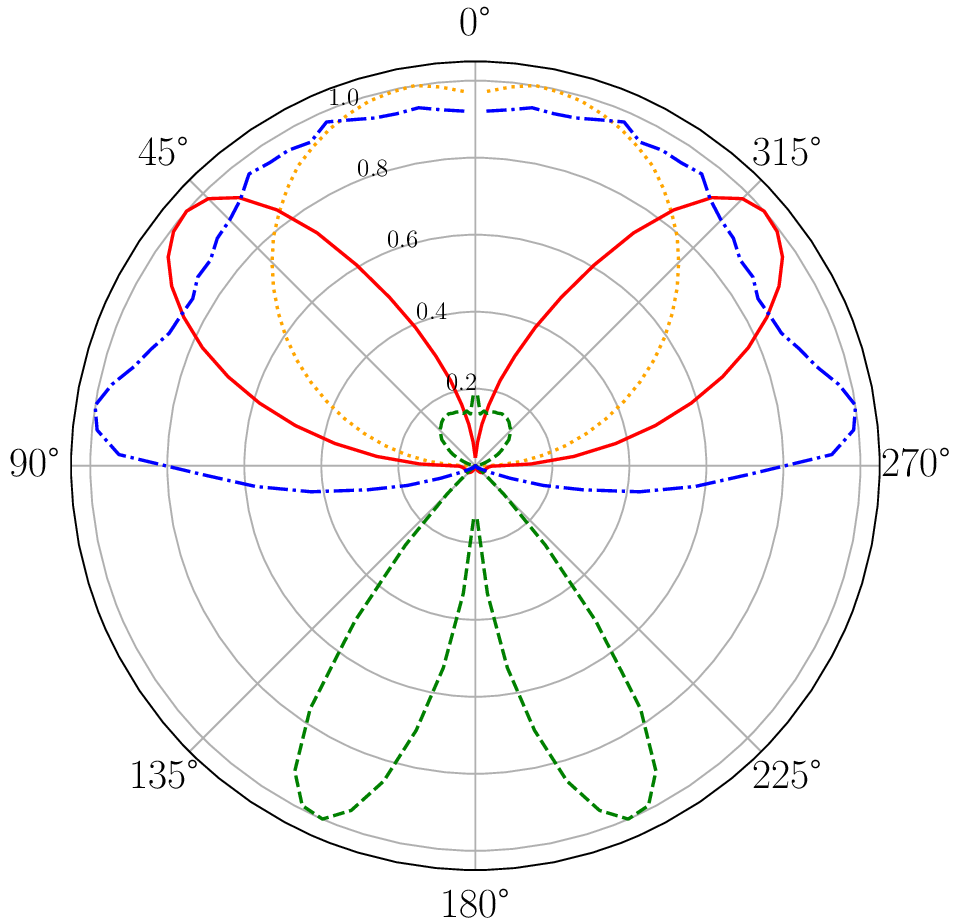}
\caption{Same as in \cref{fig:gen25}, but separately for the outgoing X-mode (upper panels) and O-mode (lower panels).
\label{fig:pol_out}}
\end{figure}  

\subsection{Polarization-resolved beaming}
\label{sec:pol_out}

In \cref{sec:boundary} we considered the influence of polarization of the \emph{incoming} radiation on the beaming of the outgoing radiation, summed over its polarization states. Let us now consider the beaming patterns separately for each of the two outgoing normal modes. 
\Cref{fig:pol_out} shows the results for the case of $\dot{M}=2.5\times10^{15}$ g s$^{-1}$, 
to be compared with \cref{fig:gen25}.
We can notice a moderate difference in the beaming of the two modes at $E=0.4\Ece$ and a significant difference at the higher energies $E\geq\Ece$,
despite the incoming radiation was assumed unpolarized. 
This difference shows that the scattering in the accretion channel substantially affects formation of the polarization of the XRP radiation
(a more extensive study and discussion of the effects of the in-channel scattering on the polarization will be given in Ref.~\cite{Markozov_26}).

\subsection{Beaming as a function of photon energy and direction}
\label{sec:3D}

\begin{figure}[ht]
\centering
\includegraphics[width=.46\columnwidth]{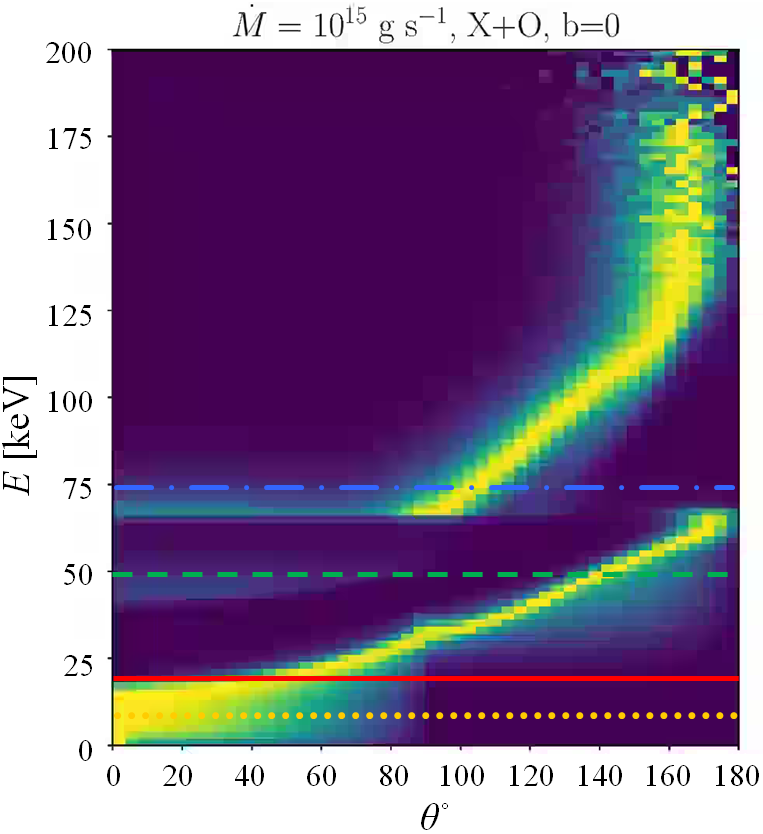}
\includegraphics[width=.52\columnwidth]{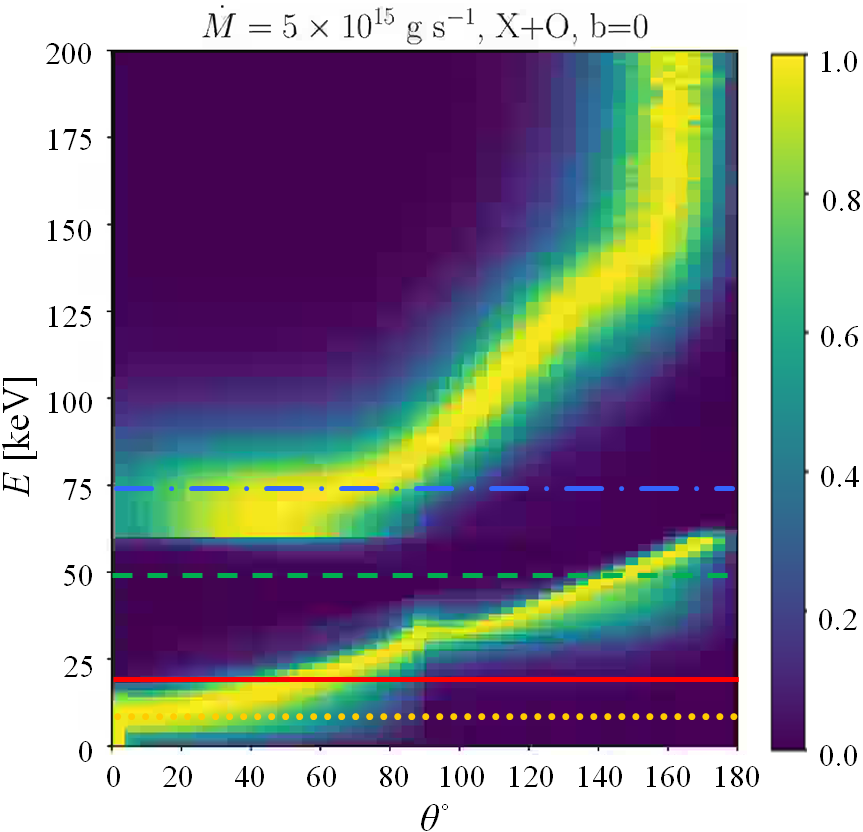}
\\
\includegraphics[width=.46\columnwidth]{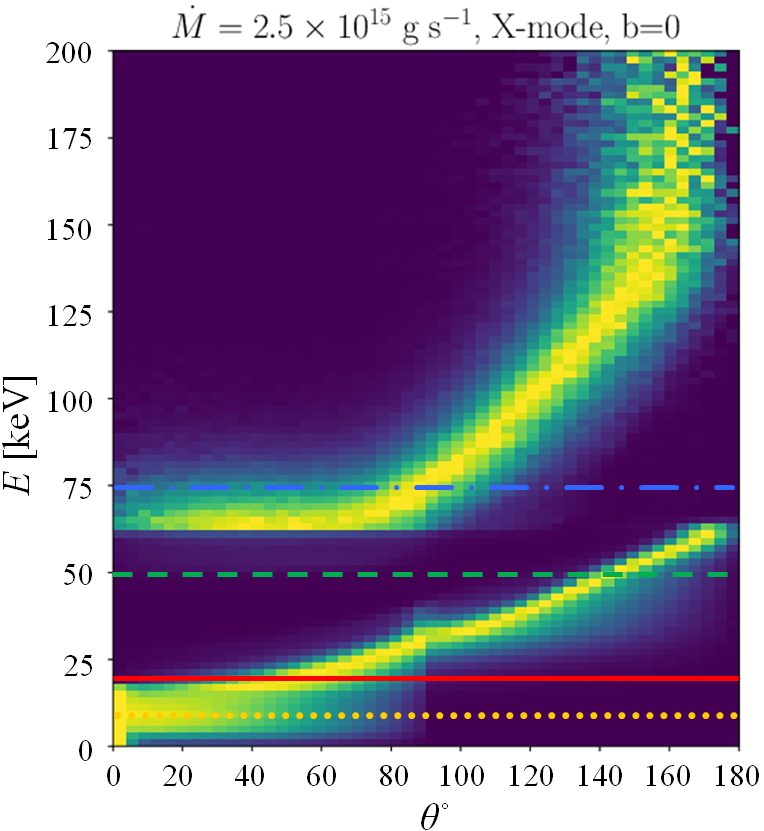}
\includegraphics[width=.52\columnwidth]{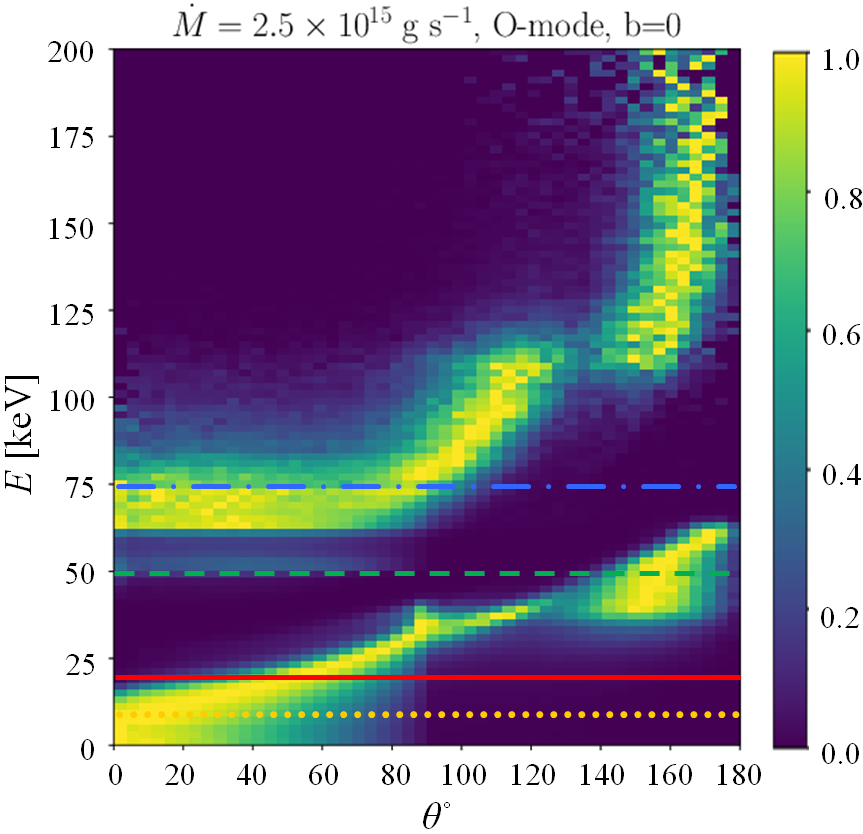}
\caption{Color maps of the normalized intensity of outgoing radiation as functions 
of the angle to the NS surface normal $\theta$ (the horizontal axis, in degrees) and photon energy $E$ (the vertical axis, in keV).
Upper panels: total intensity for both normal modes for accretion rates $\dot{M}=10^{15}$ g s$^{-1}$ (left panel) and $\dot{M}=5\times10^{15}$ g s$^{-1}$ (right panel).
Lower panels: the outgoing radiation polarized in the X-mode (left panel) and the O-mode (right panel)
at the intermediate mass accretion rate $\dot{M}=2.5\times10^{15}$ g s$^{-1}$.
The horizontal lines are drawn at the energies for which the curves of the same types
{are plotted}
in the other figures.
\label{fig:maps}}
\end{figure}

In \cref{fig:maps} the  intensity of radiation, normalized to its maximum as a function of $\theta$ at each $E$, is coded with color and shown in the $\theta$--$E$ plane. The bright yellow color corresponds to the maximum of intensity and the dark violet to its minimum. The incoming radiation was assumed unpolarized and isotropic.
This figure clarifies the origin of the differences between beaming patterns at different energies observed in \cref{fig:gen10,fig:gen25,fig:gen50,fig:pol_out}. The values of these energies are marked by the horizontal lines.

The upper panels in \cref{fig:maps} present the maps for the total intensity
{at}
two different accretion rates and the lower panels present them separately for 
the X- and O-modes at an intermediate mass accretion rate. 
We see that there are  two 
{strips of clear maxima at different energies with a minimum in between},
which drift from lower to higher $E$ with increasing $\theta$. 
{At some energies, both strips can be crossed by corresponding horizontal lines, giving a complex beaming pattern like the dot-dashed line on the right panel of \cref{fig:plasma}.}
In addition, there is a local maximum at a nearly constant energy
{close to}
$E=\Ece$, which is more clearly seen
{in the upper-left and lower-right panels of \cref{fig:maps}.}

The wide dark valleys between the pairs of  brighter strips, directed toward higher energies at larger  angles $\theta$, map  phase regions in which the  scattering optical thickness becomes greater than one. Thus scattering   processes effectively remove the original photons from the phase region of the directions and energies corresponding to the dark valleys, which  leads to the angular dependencies shown in \cref{fig:gen10,fig:gen25,fig:gen50,fig:pol_out}. 
These angular dependencies can be easily traced at each photon energy $E$ by following the corresponding horizontal lines in \cref{fig:maps}.


\subsection{Light curves}
\label{sec:lc}

The beaming patterns presented above determine the light
curves of spectral flux and polarization of an XRP. Below we present
examples of such light curves, calculated in several approximations.
First, we neglect the effects specific for NS rotation: shape flattening, general-relativistic frame dragging, Doppler boosting and relativistic
aberration \cite{Stergioulas03,PoutanenBeloborodov06,Poutanen20b}. This is a good approximation for the typical (not millisecond) X-ray
pulsars. Second, we neglect the size of the emission region compared to the NS radius. Third, we assume that the X-ray photons that have escaped from this region propagate in the empty space. 
Then each photon trajectory is a flat curve, which is accurately described by the analytical approximation of Poutanen \cite{Poutanen20a}, which 
gives the angle $\theta$ between the wave vector $\bm{k}$ and normal $\bm{n}$ to the NS surface in terms of elementary functions of the angle $\psi$ between $\bm{n}$ and the direction to a distant observer.

\begin{figure}[ht]
\includegraphics[width=.533\columnwidth]{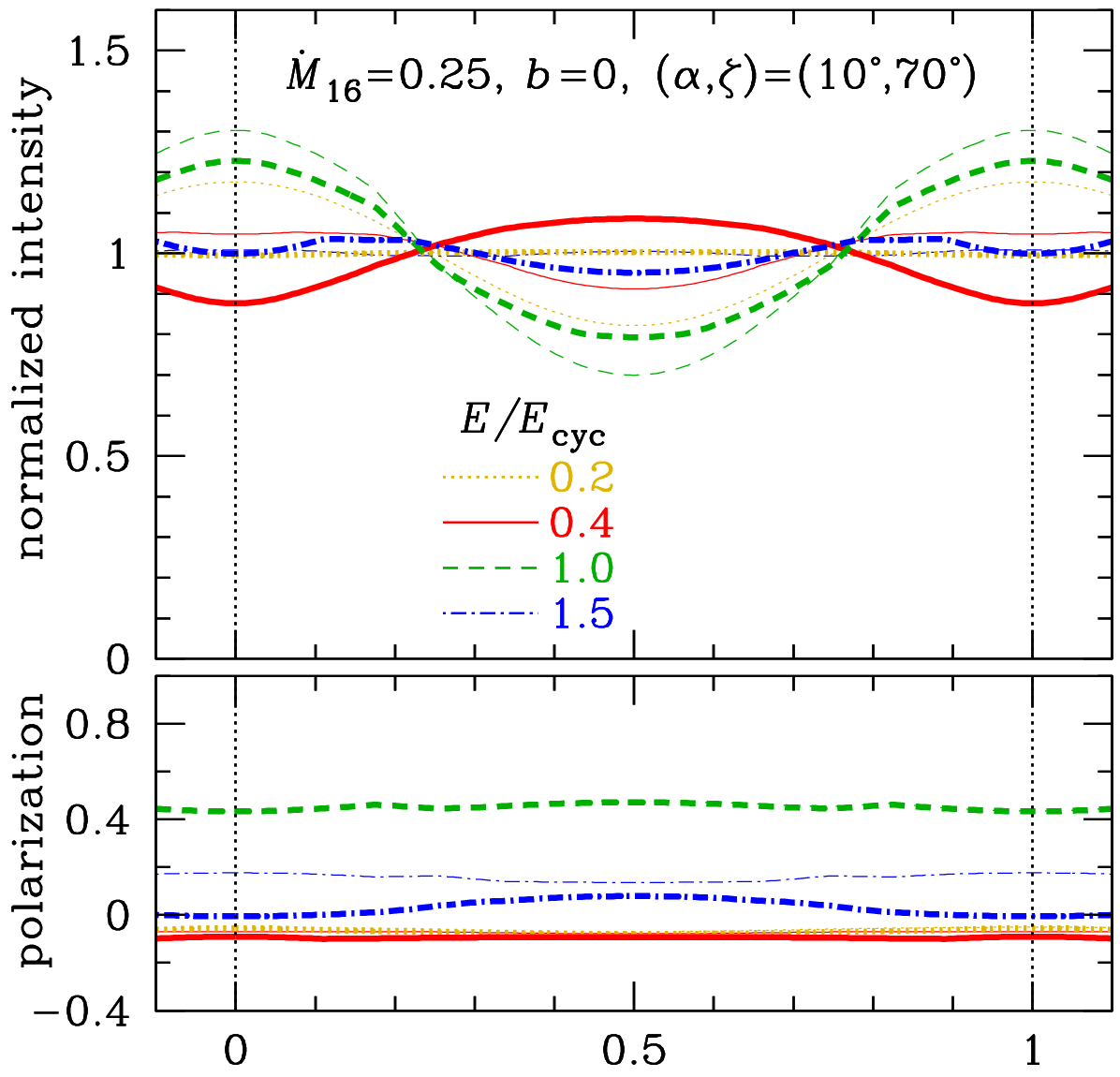}
\hfill
\includegraphics[width=.466\columnwidth]{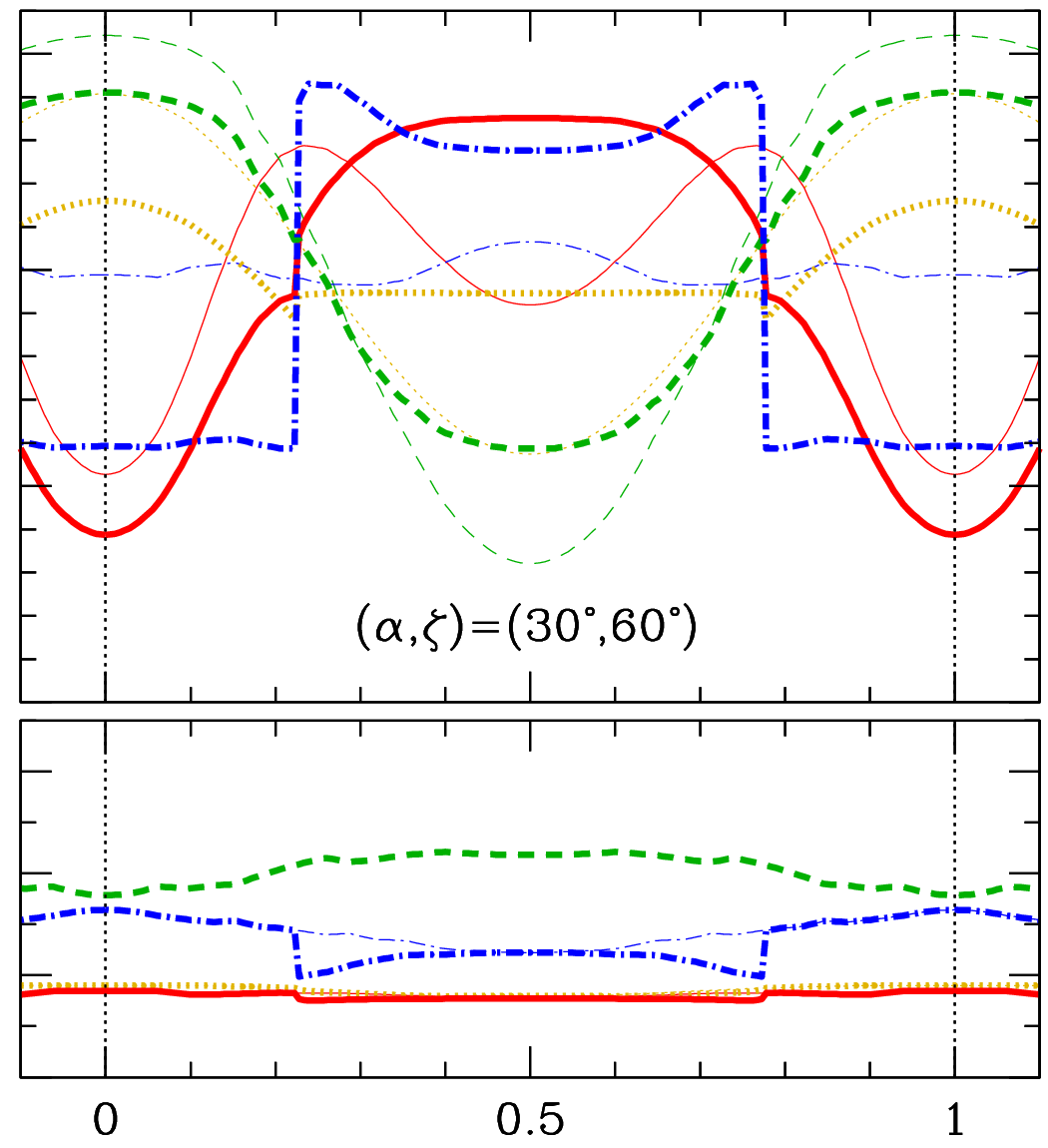}
\\
\includegraphics[width=.533\columnwidth]{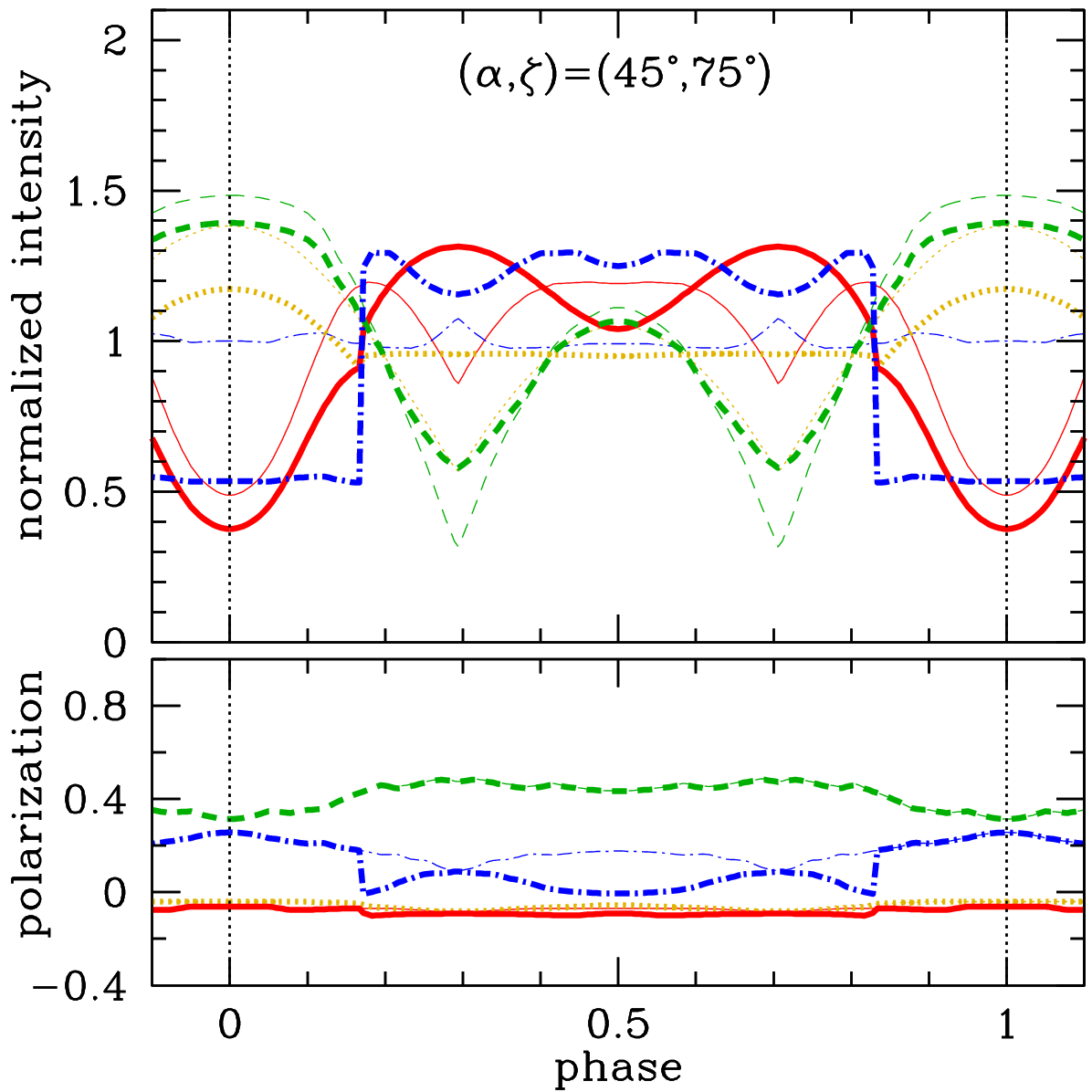}
\hfill
\includegraphics[width=.466\columnwidth]{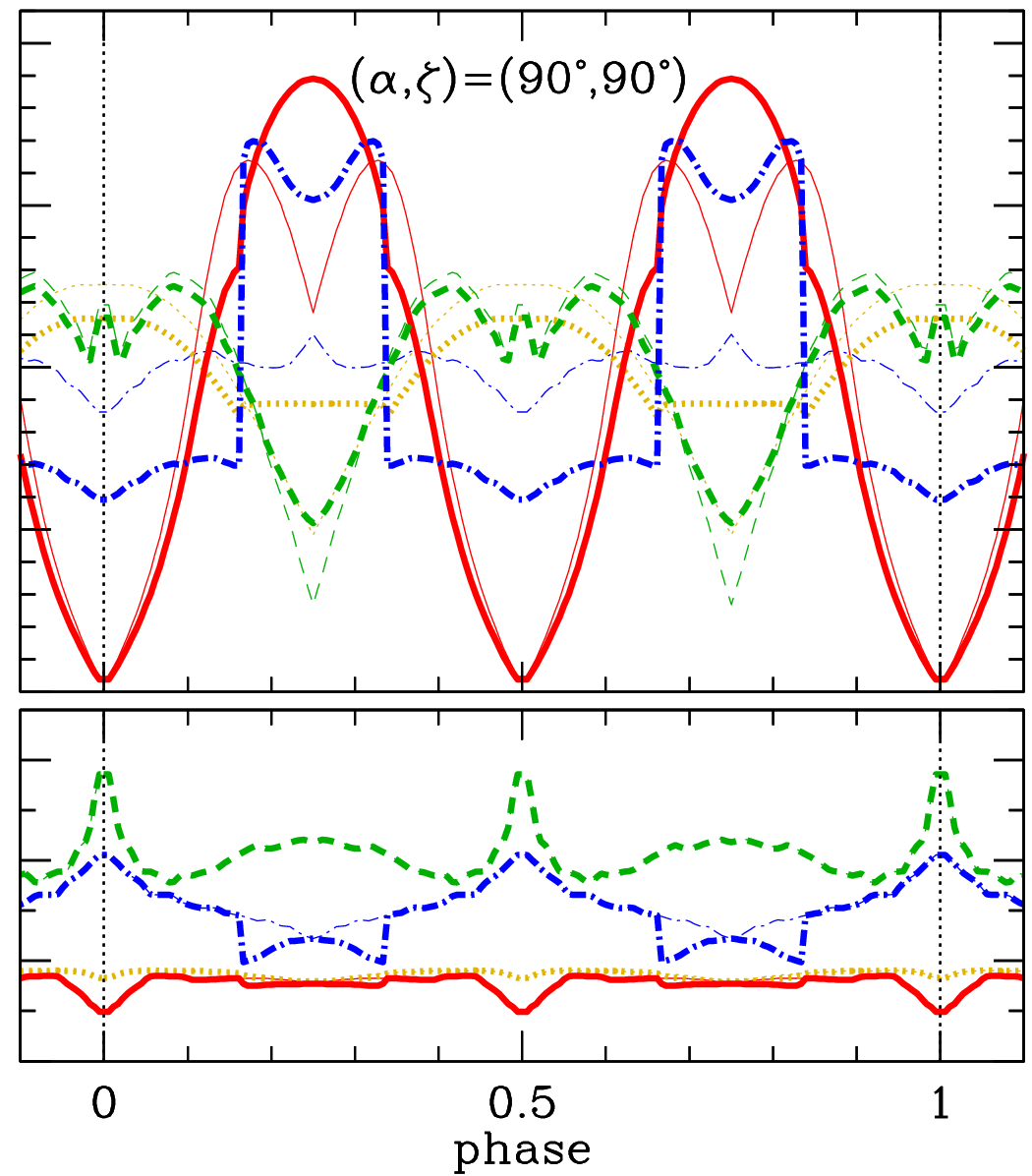}
\caption{Light curves for the intensity and polarization
degree, corresponding to the beaming patterns shown in
\cref{fig:pol_out} ($M=1.4\,M_\odot$, $R_\mathrm{NS}=
{12}$~km,
$\dot{M}=2.5\times10^{15}$ g s$^{-1}$, 
input beaming parameter $b=0$) for different combinations of the angles $\alpha$ and $\zeta$
that the spin axis makes with the magnetic axis and the line of sight.
Different line types correspond to different energies of outgoing
photons in the local NS reference frame. The thick and thin lines of the
same type show the models with two antipodal emission spots and with a
single spot, respectively.
\label{fig:lc_b0}}
\end{figure}

\begin{figure}[ht]
\includegraphics[width=.533\columnwidth]{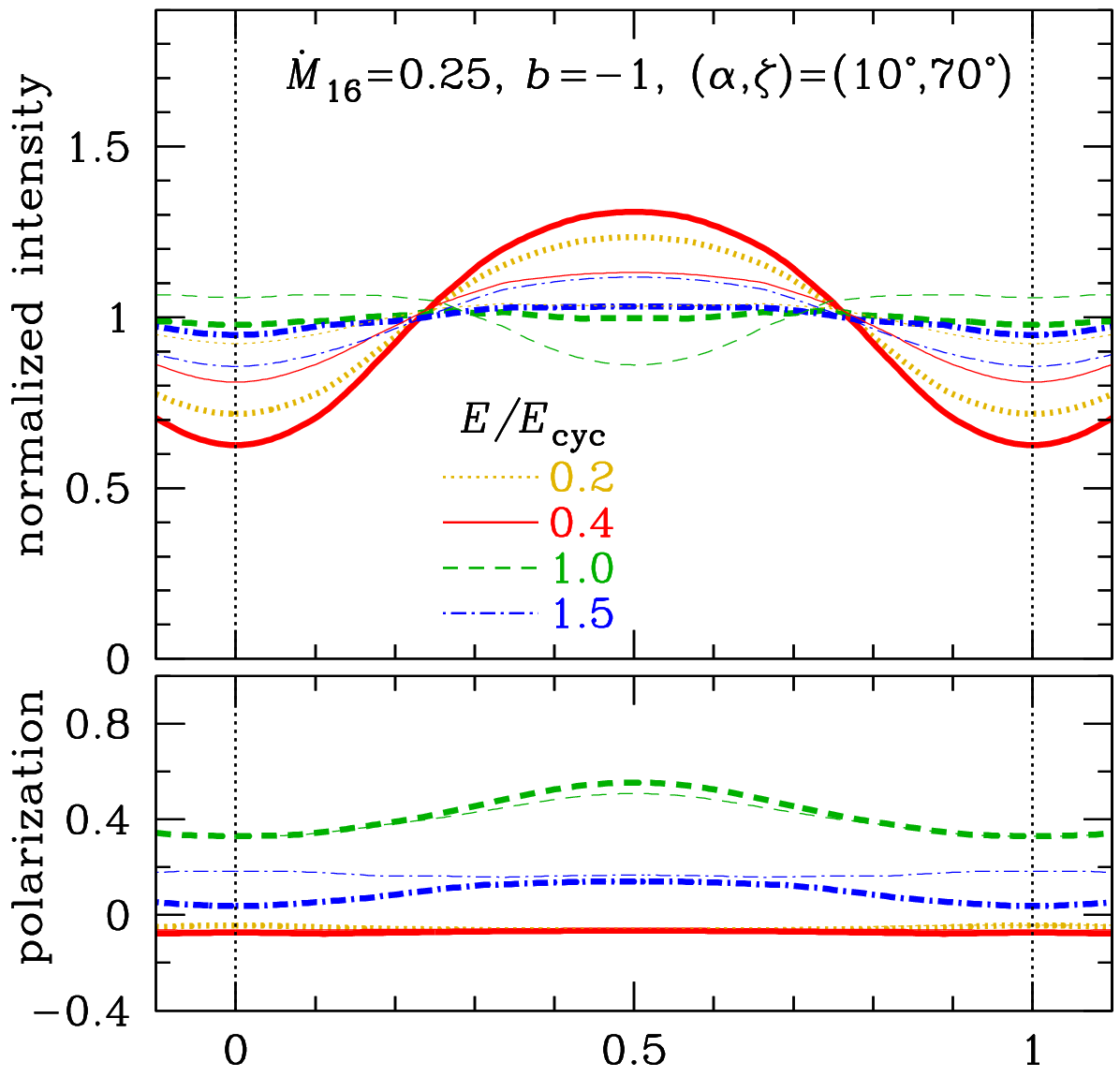}
\hfill
\includegraphics[width=.466\columnwidth]{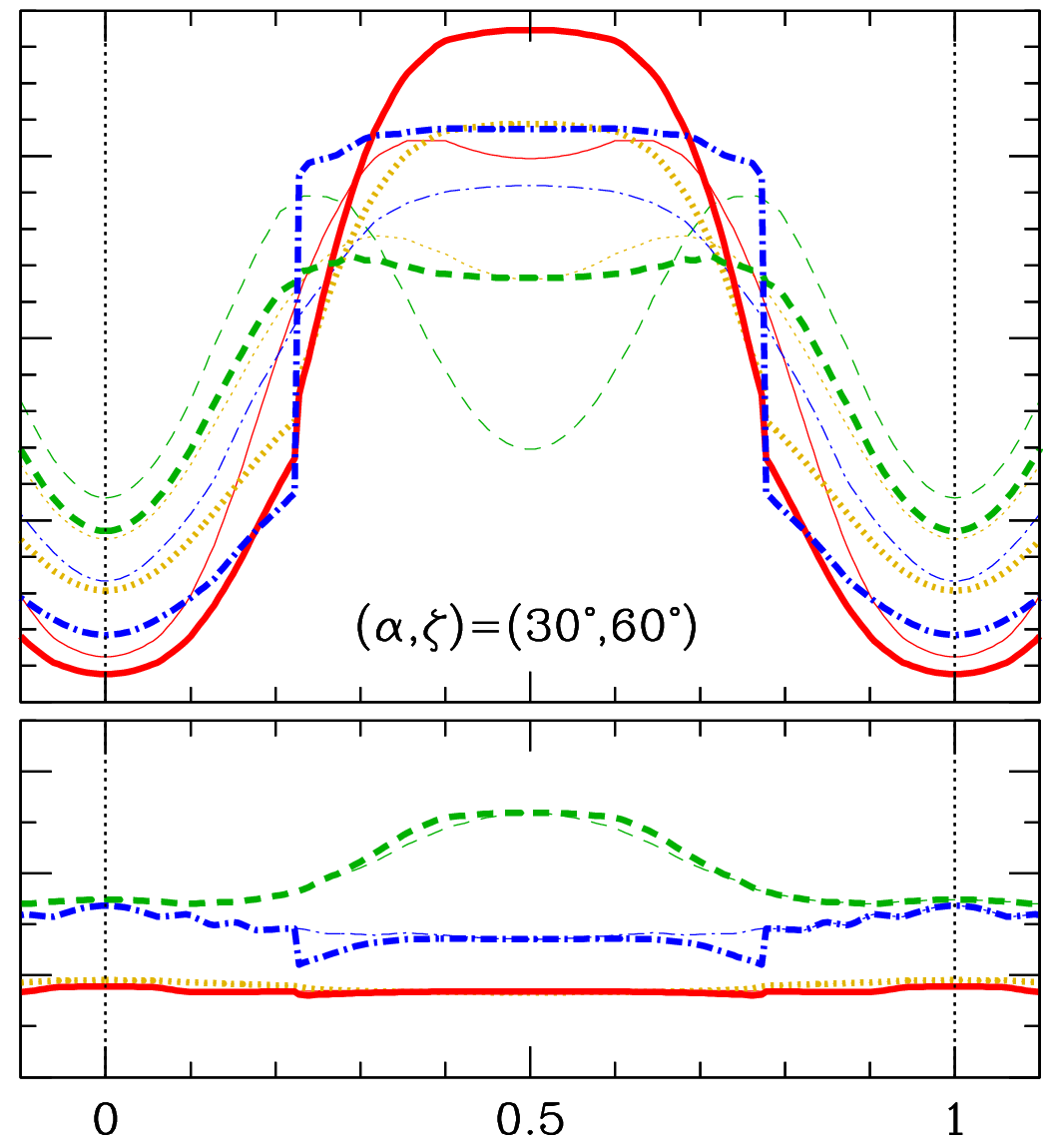}
\\
\includegraphics[width=.533\columnwidth]{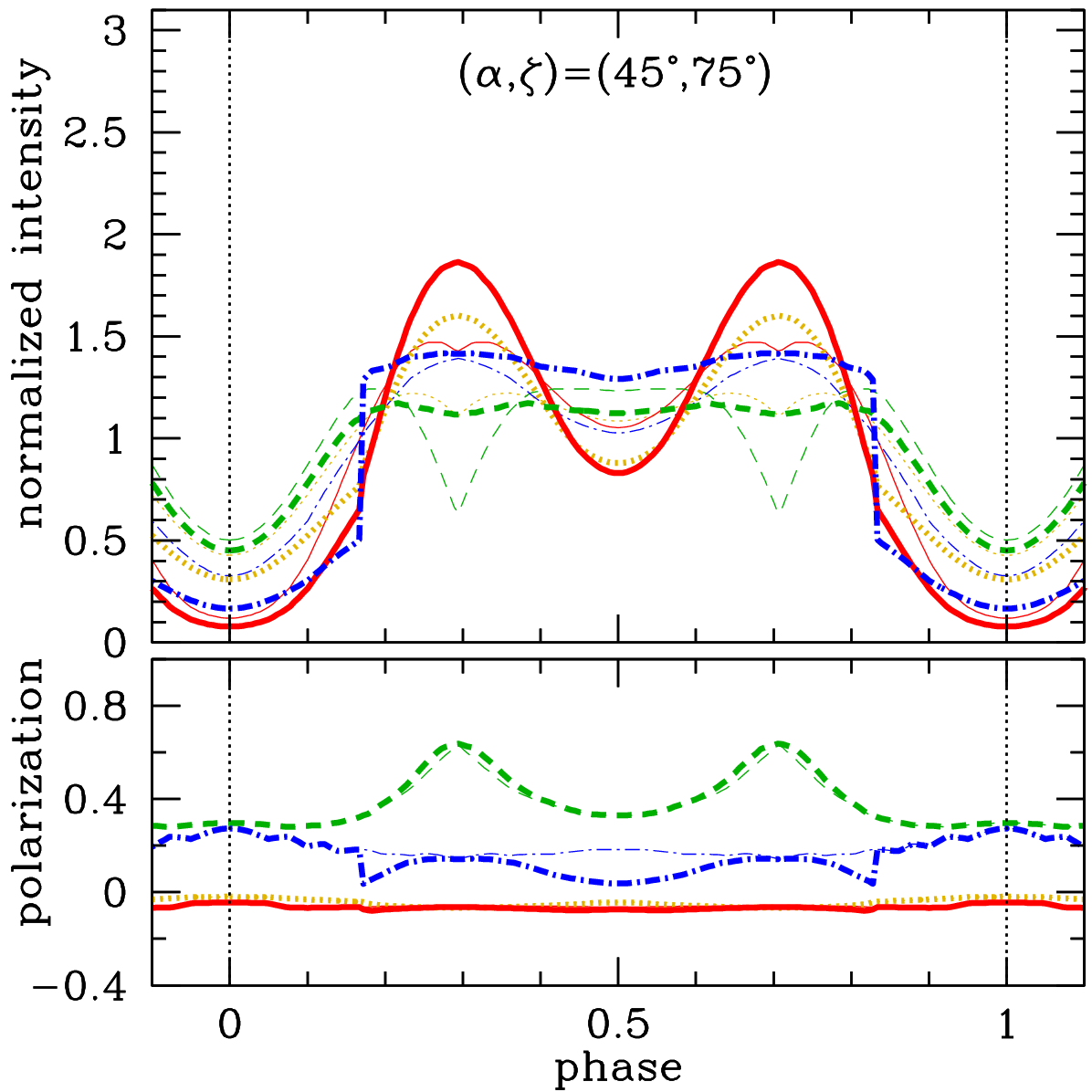}
\hfill
\includegraphics[width=.466\columnwidth]{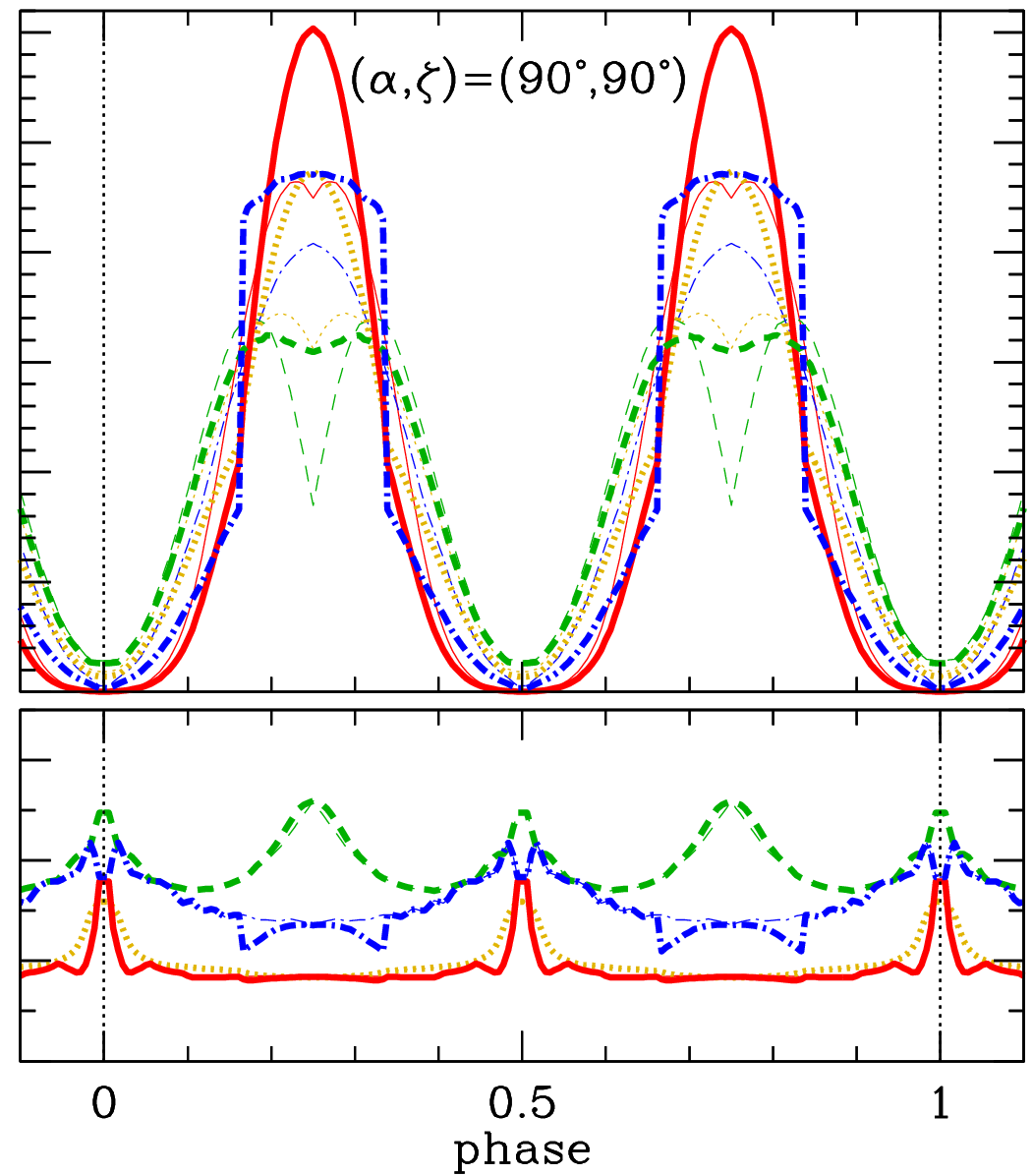}
\caption{Same as in \cref{fig:lc_b0} but for the fan-like input beaming
with parameter $b=-1$.
\label{fig:lc_bm1}}
\end{figure}

For an infinitely narrow bundle of light rays with transverse cross section $\dd S_\perp$ that reach a distant observer, the observed spectral flux density equals 
$\dd F_E^\infty = I_E^\infty \dd\Omega$, where $I_E^\infty$ is the observed specific intensity and $\dd\Omega$ is the solid angle that the
element $\dd S_\perp$ occupies in the observer's sky. 
Since the ratio $I_E/E^3$
is invariant (see, e.g., Section 22.6 in Ref.~\cite{MisnerThorneWheeler}) and the photon energy
{at infinity is decreased}
by the redshift factor $(1+z_\mathrm{g})^{-1}$, the  specific intensity in the local reference frame at the NS surface is $I_E = I_E^\infty\,(1+z_\mathrm{g})^3$. On the other hand, $\dd\Omega = \mathcal{D}\dd S_\perp/D^2$, where $D$ is the distance and  $\mathcal{D}=(1+z_\mathrm{g})^2\, \dd\cos\theta/\dd\cos\psi\approx1$ is the lensing factor, which is also accurately represented as a function of $\psi$ by the approximation of Poutanen \cite{Poutanen20a}. Thus
\beq
   F_E^\infty(\psi) = \frac{\mathcal{D}(\psi)
   Q_E(\theta(\psi))}{D^2\,(1+z_\mathrm{g})^3},
\label{F_E}
\eeq
where $Q_E=\int_{S_\perp}I_E \dd S_\perp$ is the radiant intensity at the polar angle $\theta$ in the local reference frame. 
Since the backward radiation is intercepted by the NS surface, we set $F_E^\infty(\psi) =0$ whenever $\theta(\psi) \geq 90^\circ$.

{Since we neglect the size and geometry of the emitting zone, the angle}
$\psi$ coincides with the angle between the magnetic axis of the star and the line of sight. Let $\alpha$ and $\zeta$ be the angles that the spin axis makes with the magnetic axis
and with the line of sight, respectively. Then the light curve from a single accretion channel is given by \cref{F_E} with
\beq
   \cos\psi = \sin\zeta\sin\alpha\cos(\phi - \phi_0) + \cos\zeta \cos\alpha,
\label{psi_of_phi}
\eeq
where $\phi$ is the rotational phase and $\phi_0$ is an arbitrary constant phase shift. To add an identical antipodal emitting area, one can use \cref{F_E} with $\psi$ replaced by $\psi+\pi$.

\Cref{fig:lc_b0} shows examples of light curves for the intensity and polarization degree. To produce them, we have applied \cref{F_E,psi_of_phi} to the beaming of radiant intensities of the X- and O-modes, $Q_E=Q_E^\mathrm{X}(\theta)$ and $Q_E=Q_E^\mathrm{O}(\theta)$, respectively, shown in \cref{fig:pol_out}.
{As stated above,}
we replace the real distribution of the outgoing photons over the surface of the accretion channel by a point source.
This model is relevant within the scope of the present paper, because it elucidates the clean effect of the beaming pattern, which is the main focus of study here, separated from such subtle factors as a complex geometry of the emission zone. It is expected to be acceptable for the subcritical X-ray pulsars,
{because their}
emission zones have both the radii $R_\mathrm{c}$ and heights $H$ much smaller than $R_\mathrm{NS}$
{\cite{SuleimanovLS07}}.
We have put $\phi_0=0$ in \cref{psi_of_phi}, which means that the viewing angle of the primary emission zone is minimal at $\phi=0$.
The calculations have been done at four values of the emitted photon energy $E=10$ keV, 20 keV, 50 keV and 75 keV, the same as in \cref{fig:gen10,fig:gen25,fig:gen50,fig:pol_in,fig:beamed,fig:plasma,fig:pol_out}, which correspond to the observed photon energies $E^\infty = E/(1+z_\mathrm{g})=8.1$ keV, 16.2 keV, 40.5 keV and 60.7 keV. The light curves produced by a single emission zone are shown by the thin lines, and the thick lines are given by the model of two symmetrical antipodal hot spots.

The light curves for the intensity present the sum of the flux densities $F_E^\infty$ for the X- and O-modes, which we denote respectively $F_E^\mathrm{X}(\phi)$ and $F_E^\mathrm{O}(\phi)$, calculated as functions of the rotation phase $(\phi/2\pi)$ and normalized to the mean value according to the formula
\beq
   f_E(\phi) = \frac{F_E^\mathrm{X}(\phi) + F_E^\mathrm{O}(\phi)}{
                \int_0^{2\pi} \big[ F_E^\mathrm{X}(\phi')
                 + F_E^\mathrm{O}(\phi') \big]\,\dd\phi'/2\pi}.
\eeq
We see that there can be a large variety of shapes of the light curves, depending on the geometry settings and photon energy. 
It is easy to trace the origin of the properties (minima and maxima) of each light curve in the
beaming pattern of the corresponding total radiant intensity $Q_E^\mathrm{X}(\theta)+Q_E^\mathrm{O}(\theta)$ near the NS surface, shown in \cref{fig:gen25}. 
In particular, for the lowest photon energy $E=0.2\Ece$ (the dotted orange curve) $f_E$ in \cref{fig:lc_b0} has a maximum at $\phi=0$, which corresponds to the maximum of the initial radiant intensity $Q_E$ at $\theta=0$, while for $E=0.4\Ece$ (the solid red line) $f_E$ has a minimum at $\phi=0$ and a maximum at an intermediate phase, which correspond to the fan-type beaming pattern of $Q_E$ at this $E$ value.
In \cref{fig:lc_bm1}, at contrast, the low-energy curve has a minimum at $\phi=0$, corresponding to the minimum of the initial intensity at $\theta=0$.
In the extreme case of $\alpha=\zeta=90^\circ$, each light curve in the model with two opposite hot spots displays an apparent period equal to one half of the true rotation period, because of the alternating appearance of the two identical antipodal emission zones. 
The jumps (vertical segments of the light curves) in this model occur at the phase values, where both hot spots become visible simultaneously due to the light bending (the jumps up) of where one of them becomes invisible behind the horizon (the jumps down). These jumps would be smeared over a narrow but finite phase interval, were the finite size of each emission zone taken into account. The light curves produced by a single spot trace the beaming pattern more directly, while in the two-spot model the secondary maximum on the light curve, produced by the second spot at an intermediate phase, may override the primary minimum related to the beaming -- compare, for instance, the thick and thin red lines in the case of $(\alpha,\zeta)=(30^\circ,60^\circ)$ in both \cref{fig:lc_b0,fig:lc_bm1}.

The polarization curves are calculated according to
\beq
   P_\mathrm{O-X}(\phi) = \frac{F_E^\mathrm{O}(\phi) - F_E^\mathrm{X}(\phi)}{
   F_E^\mathrm{O}(\phi) + F_E^\mathrm{X}(\phi)}.
\eeq
Taking into account that the linear polarization degree equals 
\beq
   P_L = \frac{I_\mathrm{O}-I_\mathrm{X}}{
   I_\mathrm{O}+I_\mathrm{X}}p_1,
\end{equation}
where $I_\mathrm{O}$ is the O-mode intensity, $I_\mathrm{X}$ is the X-mode intensity and $p_1$ is the linear polarization degree of a single normal mode, we see that $P_\mathrm{O-X}$ is close to the observed $P_L$, provided that $p_1\approx1$, which is indeed the case if the normal mode ellipticity is dominated by the vacuum polarization.
Although the initial radiation from the NS surface was taken unpolarized in the shown examples, the observed polarization degree is significant at some phases for sufficiently high photon energies and especially at $E=\Ece$, which highlights the effect of the resonant Compton scattering in the accretion channel at such $E$ values.

\subsection{Connection to observations}

{Although we do not perform a direct calculation of pulse profiles or the pulsed fraction (PF) as a function of photon energy, our results allow for a qualitative comparison with observational data.}

{
Observational studies of accreting X-ray pulsars indicate that both the pulse profile and the pulsed fraction depend on photon energy. 
In particular, a local minimum or, more generally, a distinct feature in PF($E$) near the cyclotron resonance energy has been reported for a number of subcritical sources (see, e.g., 
\cite{2009AstL...35..433L,2023A&A...677A.103F}). 
Such behavior is observed, for instance, in sources like V~0332+53, 4U~0115+63 and Her~X-1.
Our simulations show a qualitatively similar behavior
in most (although not all) of the considered geometries.
In the vicinity of $E \sim E_{\mathrm{cyc}}$, resonant Compton scattering redistributes photons over angles, leading to a less anisotropic emission pattern. 
As a result, the
light curve amplitude
decreases, which implies a reduction of the pulsed fraction. 
This effect is evident from
\cref{fig:lc_b0,fig:lc_bm1}, where the modulation amplitude
of the simulated light curves is smaller at the cyclotron energy $E_\mathrm{cyc}$ than at the neighboring energies ($0.4\, E_\mathrm{cyc}$ and $1.5\,E_\mathrm{cyc}$) in three of the four combinations of angles $(\alpha,\zeta)$ for the model of two symmetric hot spots, if an isotropic surface radiation is assumed (i.e., the anisotropy parameter $b=0$, the thick curves in \cref{fig:lc_b0}), and for all considered combinations of $(\alpha,\zeta)$ and both one- and two-spot models, if the surface radiation is assumed to be suppressed near the normal ($b=-1$, \cref{fig:lc_bm1}).
}

{
The fact that the cyclotron suppression of the pulsed fraction is present not for all parameter sets in our calculations
is consistent with observational results showing that the behavior of PF($E$) near the cyclotron energy is not universal. 
A quantitative analysis of
the simulated PF($E$) and its detailed comparison with observations
is beyond the scope of the present work.
}

\section{Conclusions}
\label{sec:concl}

We have studied beaming patterns and light curves of subcritical XRPs.
For typical parameters of such XRPs, we have numerically solved the radiation hydrodynamics equations for the structure of the accretion channel and performed Monte Carlo simulations of radiative transfer in this channel.
The resulting beaming patterns for the outgoing radiation at photon energies $E \geq 0.4\Ece$
differ drastically from the beaming pattern of the radiation injected into the accretion channel from the NS surface. In particular, the beaming pattern is strongly anisotropic and polarized even for an isotropic unpolarized surface emission. There is a moderate dependence of the resulting beaming pattern on the boundary conditions (the polarization and beaming of the incoming surface radiation)
at $E\geq0.4\Ece$ and a stronger dependence on the accretion rate at $E > \Ece$. We have demonstrated that the account of the vacuum polarization is necessary for a correct simulation of the beaming: were it neglected, the beaming pattern would be very different.

The calculated beaming patterns have been implemented in modelling light curves of the subcritical XRPs. Manifestations of properties of the beaming pattern have been demonstrated on examples for the dependencies of the observable flux density and polarization degree on the rotational phase of an XRP.

{Our results also provide a qualitative link to observational studies of energy-dependent pulse profiles. 
The angular redistribution of radiation caused by resonant Compton scattering near $E \sim E_{\mathrm{cyc}}$ reduces the anisotropy of the emerging emission and, consequently, the modulation amplitude of the light curves. 
This behavior is consistent with observational indications that the pulsed fraction may exhibit a local decrease or a distinct feature near the cyclotron resonance energy in some subcritical X-ray pulsars \cite{2023A&A...677A.103F}.
}

\vspace{6pt}

\authorcontributions{Conceptualization, methodology and investigation, all authors equally; software, I.M.; writing---original draft preparation, A.P.; writing---review and editing, all authors equally; visualization, I.M. and A.P. All authors have read and agreed to the published version of the manuscript.
}

\funding{The research of I.M. was supported by the Foundation for the Advancement of Theoretical Physics and Mathematics ``BASIS''.
The research of A.P. was supported by the Ministry of Science and Higher Education of the Russian Federation (agreement no. 075-15-2024-647).
}


\acknowledgments{This research was supported by the International Space Science Institute (ISSI) in Bern, through International Team project 25-657 ``Polarimetric Insights into Extreme Magnetism''.
I.M. and A.P. thank for hospitality the organizers of the international conference ``The Modern Physics of
Compact Stars and Relativistic
Gravity 2025'' in Yerevan that promoted the present work.}

\conflictsofinterest{The authors declare no conflicts of interest.} 

\begin{adjustwidth}{-\extralength}{0cm}

\reftitle{References}


\newcommand{\aap}{Astron. Astrophys.}
\newcommand{\apj}{Astrophys. J.}
\newcommand{\apjl}{Astrophys. J.}
\newcommand{\apss}{Astrophys. Space Sci.}
\newcommand{\mnras}{Mon. Not. R. Astron. Soc.}
\newcommand{\nat}{Nature}
\newcommand{\pasj}{Publ. Astron. Soc. Japan}
\newcommand{\prd}{Phys. Rev. D}
\newcommand{\sovast}{Sov. Astron.}

\bibliography{allbib}

\end{adjustwidth}
\end{document}